\documentclass[12pt]{article}

\usepackage{a4}
\usepackage{a4wide}

\usepackage{amsmath}
\usepackage{amscd}
\usepackage{amsfonts}
\usepackage{amssymb}
\usepackage{mathrsfs}
\usepackage{fancybox} 
\usepackage{enumerate}

\usepackage{cite}
\usepackage{bm}
\usepackage{amscd}
\usepackage{graphicx}
\usepackage[dvips]{color}
\usepackage{epsfig}

\usepackage{braketmod}

\makeatletter

\@addtoreset{equation}{section}
\makeatother
\usepackage{amsthm} 

\def\tr{\mathrm{tr}}
\def\Tr{\mathrm{Tr}}

\def\det{\mathrm{det}}
\def\det{\mathrm{det}}

\def\adj{\mathrm{adj}}

\def\half{{1\over2}}
\def\sgn{\mathrm{sgn}}

\def\gh{\mathrm{gh}}

\newcommand{\wt}{\widetilde}

\def\={\stackrel{\bullet}{=}}

\def\({\left(}
\def\){\right)}
\def\[{\left[}
\def\]{\right]}

\def\cD{{\cal D}}

\def\cN{{\cal N}}
\def\cO{{\cal O}}

\def\mbf{\mathbf}

\def\mf {\mathfrak }
\def\mbb {\mathbb }

\def\ms {\mathsf }
\def\mr {\mathrm }

\def \be {\begin{equation}}
\def \ee {\end{equation}}
\def \beqa {\begin{eqnarray}}
\def \eeqa {\end{eqnarray}}
\def \beal#1 {\begin{align}#1\end{align}}
\def \bes#1 {\begin{equation}\begin{split}#1\end{split}\end{equation}}
\def \nn {\notag\\}

\def\sla#1{\not\!\!#1}

\def\bra#1{\langle #1 |}

\def\ket#1{|#1 \rangle}
\def\kket#1{|#1 \rangle\rangle}
\def\aver#1{\left\langle #1 \right\rangle}

\def\pmat#1{\begin{pmatrix}#1 \end{pmatrix} }

\global\long\def\bra#1{\Bra{#1}}%

\global\long\def\ket#1{\Ket{#1}}%

\global\long\def\kket#1{\Kket{#1}}%

\global\long\def\braket#1{\Braket{#1}}%

\global\long\def\bbraket#1{\Bbraket{#1}}%

\global\long\def\brakket#1{\Brakket{#1}}%

\global\long\def\bbrakket#1{\Bbrakket{#1}}%

\global\long\def\hsum{\sum_{h_{1},\ldots,h_{N}=0}^{n-1}}%

\makeatletter

\@addtoreset{equation}{section}
\makeatother

\begin{document}

\begin{titlepage}
\title{
\vspace{-2cm}
\begin{flushright}
\normalsize{ 
YITP-21-83 \\ 
}
\end{flushright}
\vspace{1.5cm}
Topological phase, spin Chern-Simons theory and \\
level rank duality on lens space
\vspace{1.cm}
}
\author{
Naotaka Kubo\thanks{naotaka.kubo[at]yukawa.kyoto-u.ac.jp}
\hspace{5mm}and\hspace{5mm}
Shuichi Yokoyama\thanks{shuichi.yokoyama[at]yukawa.kyoto-u.ac.jp},\; 
\\[25pt] 
$^{*}$$^{\dagger}$\,{\normalsize\it Center for Gravitational Physics,} \\
{\normalsize\it Yukawa Institute for Theoretical Physics, Kyoto University,}\\
{\normalsize\it Kitashirakawa Oiwake-cho, Sakyo-Ku, Kyoto 606-8502, Japan}
}

\date{}

\maketitle

\thispagestyle{empty}

\begin{abstract}
\vspace{0.3cm}
\normalsize

We study a method to compute a topological phase factor of partition function for pure Chern-Simons theory incorporating the supersymmetric localization.
We develop a regularization preserving supersymmetry and the topological phase appears as a result of the supersymmetric regularization. 
Applying this method to pure Chern-Simons theory on lens space we compute the background dependent phase factor coming from the Chern-Simons term. 
We confirm that the partition function computed in this method enjoys a couple of level rank dualities including the one recently proposed in arXiv:1607.07457 for all ranks and levels within our numerical calculation. 
We also present a phase factor with which the lens space partition function exhibits the perfect match between any level rank dual pair including the total phase.

\end{abstract}
\end{titlepage}
\tableofcontents

\section{Introduction}
\label{Intro} 

Topological phase has recently been recognized to play a physical role to unravel rich structure in quantum field theory. 
Such a phase factor typically arises as that of partition function of topological quantum field theory (TQFT). Even if TQFT has no dynamical degrees of freedom, the phase factor becomes physical for spin TQFT by coupling it to a bosonic theory with a suitable discrete symmetry, which turns the bosonic theory into a fermionic one without adding new dynamical degrees of freedom. 

Topological property of phase is closely related to the Atiyah-Patodi-Singer index theorem. 
This was first explicitly shown in \cite{Witten:1988hf} by explaining how to compute partition function for pure Chern-Simons theory defined by path-integral as topological invariant on three manifold. 
Computation thereof needs specifying gauge fixing and regularization, which inevitably accompanies with a choice of metric and breaks its manifest topological nature. The solution prescribed in \cite{Witten:1988hf} to restore the topological property is to add a gravitational Chern-Simons term as a finite counter term so as to convert the dependence on a choice of section on the tangent bundle into that of a trivialization of the vector bundle employing the index theorem. 
Trivialization of the tangent bundle is not unique and is specified by an integer, which should be suitably chosen for the partition function to be real. 

Various techniques to exactly compute partition function of pure Chern-Simons theory on a compact orientable three manifold have been developed \cite{1992CMaPh.147..563J,Blau:1993tv,Marino:2002fk,Beasley:2005vf,Blau:2006gh,Kallen:2011ny}. (See also \cite{Blau:2013oha,Blau:2018cad}.)
In particular for this TQFT the supersymmetric localization can be applied by completing the gauge field into the $\cN=2$ vector multiplet and supersymmetrizing the action \cite{Kapustin:2009kz}.
The supersymmetric localization technique has an advantage to extend exact results to include the contribution of supersymmetric matter. 

An aim of this paper is to study how the topological property is restored when we compute the partition function for pure Chern-Simons theory on lens space adopting the supersymmetric localization and in particular to calculate its topological phase factor as a result of a regularization preserving supersymmetry, which we refer to as supersymmetric regularization. (See also \cite{2017JHEP...12..081O} in two dimensional case.)
In fact there seems to remain an issue in the lens space partition function for pure Chern-Simons theory: two different results are known for the phase factor arising from the Chern-Simons term on lens space \cite{2002math......9403H,Brini:2008ik,Gang:2009qdj,Imamura:2012rq,Imamura:2013qxa,Alday:2013lba}. 
The study of abelian Chern-Simons theory from two dimensional conformal field theory  suggests that the mismatch arises due to an incognizant choice of spin structure in the derivation \cite{Okuda:2020fyl}. This is anticipated from the fact that the mismatch disappears for even Chern-Simons level, and Chern-Simons theory becomes a spin TQFT for odd case \cite{Dijkgraaf:1989pz}. (See also \cite{Belov:2005ze}.)

Similar subtleties were known for the level-rank duality. (See  \cite{Nakanishi:1990hj,Kuniba:1990im,Naculich:1990hg,Mlawer:1990uv} for some basic literature.)
One of them is to impose some restriction to the level of WZNW model to show the level rank duality \cite{Naculich:2005tn,Naculich:2007nc} in order for primary fields to have integral dimension \cite{Gepner:1992kx}. 
If the theory is out of the restriction, then operators in the chiral algebra have half-integer dimension and modular matrices are not quite well-behaved. 
This is a signal for the theory to be spin and the resolution for the subtleties due to the case of the theory to be spin has been proposed in \cite{Hsin:2016blu}. 
It has been argued that the addition of a gravitational sector consisting of a spin half object as well as suitable background fields resolves the known issues. (See also \cite{Schnitzer:2021rgv,Schnitzer:2021txf} for recent study of the duality.)

Another aim of this paper is to provide evidence for the well-known level rank duality as well as the one recently proposed in \cite{Hsin:2016blu} for an arbitrary rank and level employing the partition function on lens space computed by the supersymmetric localization and regularization. 
Using lens space is important to test the duality involving spin theories because it admits two spin structures for an even number of the fundamental group while the unique one for an odd number \cite{Freed:1991wd}.
Supersymmetric regularization newly developed in this paper provides an alternative way to compute the phase factor coming from the Chern-Simons term and resolve the issue mentioned above. 
We confirm that the resulting partition function exhibits the level rank dualities mentioned above up to an overall phase. 
This means that the derived partition function depends on the spin structure, which is correctly chosen for the level rank duality to hold.%
\footnote{We have checked that the partition function with the other different phase factor for the Chern-Simons term mentioned above does not pass the test of the level rank duality. }
This method also provides a way to fix the overall topological phase of the theory by computing a finite counter term consisting of the supersymmetrized gravitational Chern-Simons term.
In this paper, instead of specifying a correct finite counter term, a certain phase factor is presented, with which the partition function enjoy the dualities mentioned above including the total phase. 
This provides strong evidence for newly proposed level rank duality in \cite{Hsin:2016blu}.
We expect that this phase factor is a desired topological phase and obtained as a result of adding a correct counter term for the supersymmetric regularization. 

The rest of this paper is organized as follows. 
In section~\ref{sec:Localizatioun}, we perform the supersymmetric localization for squashed lens space by employing the supersymmetric regularization and obtain matrix models. 
In section~\ref{sec:MatrixModel}, we show some properties of the matrix model of ${\rm U}(N)$ and ${\rm SU}(N)$ gauge group.
Especially, we apply a technique similar to the Fermi gas formalism to the matrix model, which evaluate the matrix integral over the Cartan of the gauge group.
In section~\ref{sec:Duality}, we find phases of the matrix models which make some level-rank dualities complete.
Finally, in section~\ref{Discussion}, we give some comments on our results and some future directions.
In appendix~\ref{gcs}, we show the computation of gravitational Chern-Simons term.
In appendix~\ref{eta-invariant}, we compute the eta invariant on lens space under non-trivial holonomy background.
In appendix~\ref{sec:DecomLemma}, we provide a formula which is useful for relating the matrix models of ${\rm U}(N)$ and ${\rm SU}(N)$ gauge group.

\section{Supersymmetric localization and regularization} \label{sec:Localizatioun} 

\footnote{ 
This section contains a number of technicalities to reach our final answer. A reader who is interested in our final answer of the partition function to exhibit the level rank duality can skip this section and start to read from the next section.  
}
In this section we illustrate how to compute the partition function for pure Chern-Simon theory on a squashed lens space in the form of path integral
\be 
Z[L_b(n,1), G_{k_0}] = \[ \int \cD A e^{{i k_0\over 4\pi} \int_{L_b(n,1)} \tr( A \wedge dA + {2 \over 3} A\wedge A \wedge A)  } \]_{\rm reg}, 
\label{defCSpf} 
\ee
where $G$ is a simple Lie group describing gauge symmetry and $k_0$ is an integer called the (unrenormalized) Chern-Simons level. The notation $\[ X \]_{\rm reg}$ means that the quantity $X$ inside the bracket may be divergent so as to be suitably regularized or may be added by counter terms suitably. $\mf g$ denotes the associated Lie algebra. 

In this paper we consider two types of squashed lens space. 
One is based on the squashed three sphere endowed with the metric $ds^2 = \ell_1^2 + \ell_2^2 + (\frac{1+b^2}{2b})^2\ell_3^2$, where $\ell_i$ is the left-invariant 1-form for the unit round sphere.
The other type is based on a squashed three sphere $b^2(x_1^2+x_2^2)+ b^{-2}(x_3^2+x_4^2)=1$.
See appendix~\ref{gcs} for more detail. 
The first squashing breaks the isometry from $so(4)\simeq su(2)_L\times su(2)_R$ to  $su(2)_L\times u(1)_R$, while the second one  $u(1)_L\times u(1)_R$. 
Consider the discrete subgroup of order $|n|$ in $u(1)_R$, which generates translation for the Hopf fiber direction when the sphere is realized by the circle bundle over two sphere.
Taking the quotient for the squashed three spheres by this discrete subgroup yields a squashed lens space considered in this paper. 

Since for negative $n$ the lens space $L(n,1)$ is diffeomorphic to $L(|n|,1)$ with orientation reversed, we assume $n$ is positive for simple presentation in what follows.\footnote{ Note that the partition function on $L(n,1)$ is obtained from that on $L(|n|,1)$ by analytic continuation with respect to $n$ for abelian case \cite{Okuda:2020fyl}.}     

\subsection{For squashed three-sphere} 
\label{squashedS3} 

\footnote{ 
We owe the analysis in this subsection and the following one to discussions with T.~Okuda and his unpublished note. We would like to thank him for valuable discussions. }
We first illustrate the supersymmetric localization in the case with $n=1$, or a squashed three sphere and fix a supersymmetric regularization to make the partition function \eqref{defCSpf} finite. Once we fix such a regularization, it can be used also for the case of a squashed lens space since the local structure remains intact by taking the quotient. 
The squashed three spheres considered in this paper are known to preserve $\cN=2$ supersymmetry with suitable background gauge field for R-symmetry turned on \cite{Hama:2011ea,Imamura:2011wg}. See also \cite{Pasquetti:2011fj,Dimofte:2011gm,Nian:2013qwa}.
Therefore, as mentioned in the introduction, we add some auxiliary fields so as to make the action supersymmetric and apply the supersymmetric localization to this TQFT. 

A key point in the technique is that a special term called a $Q$-exact term can be added into the action without changing the partition function. By choosing a $Q$-exact term so as to contain a free kinetic term of the supersymmetric multiplet and making its coupling constant infinitely large, the partition function defined by path-integral can be evaluated in the weak coupling limit so that the saddle point approximation becomes exact. 
The saddle points are given by the vacuum expectation values of the adjoint scalar in the vector-multiplet\footnote{Here we considered the so-called Coulomb branch localization. The saddle points are different if one uses a different Q-exact term for localization such as the so-called Higgs branch localization.} which are valued in the Cartan subalgebra of $\mf g$ and parametrized by a $r$-dimensional vector $\vec\phi=(\phi_1,\cdots,\phi_r)$ with $r$ the rank of the Lie algebra. 
Then the partition function \eqref{defCSpf} reduces to 
\beal{
Z[\mbf S_b^3, G_{k_0}] 
=&{1 \over |W|} \int_{\mbf R^r} d^r\vec\phi\, e^{-S_0 } \[ Z^{\text{1-loop}}  \]_{\rm reg} ,
}
where $|W|$ is the number of the Weyl group of the Lie algebra, $\Delta$ is the set of the roots, $S_0$ is the classical contribution given by 
\be 
S_0=\pi i k_0 |\vec\phi|^2,
\ee
and $Z^{\text{1-loop}}$ is the 1-loop determinant from the vector multiplet computed as  
\be 
Z^{\text{1-loop}}= \prod_{\vec\alpha \in \Delta}\wt s_{b}(\vec\alpha\cdot\vec\phi-{b+ b^{-1}\over2i}),
\ee 
with 
\beal{
\wt s_b(\sigma) 
=& \prod_{p,q \in \mbb Z_{\geq0} } {b(p+\half) + b^{-1} (q+\half)  -i\sigma \over b(p+\half) + b^{-1} (q+\half) + i\sigma }. 
\label{wtsb}
}

The infinite product in the partition function with \eqref{wtsb} arises as a result of the Gaussian integration for the infinite modes. 
We wish to regularize this without breaking supersymmetry, which is newly developed in this paper.\footnote{In this paper we only consider pure Chern-Simons theory, but the technique developed in this paper can be applied to any $\cN=2$ gauge theory.}  
To this end we introduce a large enough odd number of {\it chiral} multiplets in the adjoint representation for the gauge algebra as a set of regulators and counter-terms.%
\footnote{ 
The reason why chiral multiplets work as regulators of a vector multiplet may be because in three dimensions dualizing a vector field gives a scalar field. 
} 
The total number is not important. 
We label each chiral multiplet by $\wt a=1,\cdots, N_{\gh}$. For $\wt a$ even, we assign the usual statistics for each component field in the sense to obey the spin-statistics theorem. For $\wt a$ odd, we assign the opposite one. We denote the statistics of $\wt a$-th chiral multiplet by $\epsilon_{\wt a}=(-1)^{\wt a}$. Note that 
\be 
\sum_{\wt a=1}^{N_\gh} \epsilon_{\wt a}= -1.
\ee
Introduction of the chiral multiplets accompanies with the abelian flavor symmetry, which we denote by $U(1)_{\rm gh}$. We denote the $U(1)_\gh$ charge of $\wt a$-th chiral multiplet by $c_{\wt a}$. We assume that all charges are non-vanishing and satisfy the following relations. 
\be 
\sum_{\wt a=1}^{N_\gh} \epsilon_{\wt a} c_{\wt a}= 
\sum_{\wt a=1}^{N_\gh} \epsilon_{\wt a} \sgn(c_{\wt a})c_{\wt a}= 
\sum_{\wt a=1}^{N_\gh} \epsilon_{\wt a} \sgn(c_{\wt a}) c_{\wt a}^2= 0.
\label{flavorvectorghost}
\ee
Later we will see that \eqref{flavorvectorghost} is requested to cancel the divergence. 
We also assign the following relation. 
\be 
\sum_{\wt a=1}^{N_\gh} \epsilon_{\wt a} \sgn(c_{\wt a})=-\sgn(k_0). 
\label{flavorvectorghost2}
\ee
This relation will be required for the correct level shift as seen later.

In addition, for each $\wt a$-th chiral multiplet, we add $r$ gauge singlet chiral multiplets with the same charges but the opposite statistics.
For reader's convenience, we summarize the added chiral multiplets and their charge assignment in Table~\ref{ChargeReg}.
\begin{table}[htb]
\begin{center}
 \begin{tabular}{|c|ccc|c|c}
  \hline 
 Symmetry & $G$ & $U(1)_{\rm gh}$ & $U(1)_R$ & statistics \\
  \hline
Vector multiplet & adj &$0$ & 1 & 1 \\ 
$\wt a$-th chiral multiplet & adj & $c_{\wt a}$  & 1 & $\epsilon_{\wt a}$ \\ 
$r$ chiral multiplets for each $\wt a$ & $\mbf 1$ & $c_{\wt a}$ & 1 & $-\epsilon_{\wt a}$ \\ 
  \hline 
  \end{tabular}
 \caption{Symmetries and their representation of a vector multiplet and those in regulator chiral multiplets are shown. The charges written in the $U(1)_R$ column is that for the fermionic component of the super multiplet. }
 \label{ChargeReg}
\end{center}
\end{table}

We then introduce the mass of the added regulator chiral multiplets without breaking supersymmetry by turning on a background vector multiplet for the U(1)$_\gh$ symmetry. 
We denote the vacuum expectation value of the scalar component of the background vector multiplet by $\Lambda$, which plays a role of the cut-off parameter. 
Note that due to the assumption that the regulator chiral multiplets are all charged under $U(1)_{\rm gh}$ and are all massive. 

On this preparation we redo the localization procedure by adding $Q$-exact terms constituted by the regulator chiral multiplets into the action.
Then the 1-loop determinant changes as  
\beal{
Z^{\text{1-loop}}_\Lambda=& \prod_{\vec\alpha \in \Delta}\wt s_{b}(\vec\alpha\cdot\vec\phi-{Q\over2i})  \prod_{\wt a=1}^{N_\gh} \( \prod_{\vec\rho \in R_{\adj}}  \wt s_{b}(\vec\rho\cdot\vec\phi+ c_{\wt a}\Lambda - {Q\over2i} )^{\epsilon_{\wt a}} \cdot (\wt s_{b}(c_{\wt a}\Lambda - {Q\over2i} )^{-\epsilon_{\wt a}})^r \)\nn
=& \prod_{\vec\alpha \in \Delta}\left\{\wt s_{b}(\vec\alpha\cdot\vec\phi-{Q\over2i}) \prod_{\wt a=1}^{N_\gh} \wt s_{b}(\vec\alpha\cdot\vec\phi+ c_{\wt a}\Lambda - {Q\over2i} )^{\epsilon_{\wt a}} \right\} , 
}
where $R_{\adj}$ represents the set of the weights of the adjoint representation\footnote{$R_{\adj}$ contains zero vectors corresponding to the Cartan generators while the set of the roots $\Delta$ does not. } and $Q:=b+ b^{-1}$.
Then the infinite product inside the bracket is computed as follows. 
\beal{
&\wt s_{b}(\vec\alpha\cdot\vec\phi-{Q\over2i}) \prod_{\wt a=1}^{N_\gh} \wt s_{b}(\vec\alpha\cdot\vec\phi+ c_{\wt a}\Lambda - {Q\over2i})^{\epsilon_{\wt a}} \nn 
=& \exp \bigg[  \sum_{p,q \in \mbb Z_{\geq0} } \bigg\{ \log {b(p+\half) + b^{-1} (q+\half)  -i(\vec\alpha\cdot\vec\phi-{Q\over2i}) \over b(p+\half) + b^{-1} (q+\half) + i(\vec\alpha\cdot\vec\phi-{Q\over2i}) } \nn
&~~~~~+\sum_{\wt a=1}^{N_\gh} \epsilon_{\wt a} \log {b(p+\half) + b^{-1} (q+\half)  -i(\vec\alpha\cdot\vec\phi+ c_{\wt a}\Lambda - {Q\over2i}) \over b(p+\half) + b^{-1} (q+\half) + i(\vec\alpha\cdot\vec\phi+ c_{\wt a}\Lambda - {Q\over2i}) }  \bigg\} \bigg] \nn
=&\exp\[ \int_0^\infty {dt \over t} \bigg\{ { -i \sin{t(\vec\alpha\cdot\vec\phi-{Q\over2i})} \over 2\sinh {bt \over 2} \sinh {b^{-1}t \over 2} }  +\sum_{\wt a=1}^{N_\gh} \epsilon_{\wt a} { -i \sin{t(\vec\alpha\cdot\vec\phi+ c_{\wt a}\Lambda - {Q\over2i})} \over 2\sinh {bt \over 2} \sinh {b^{-1}t \over 2} }  \bigg\} \] \nn
=&s_{b}(\vec\alpha\cdot\vec\phi-{Q\over2i}) \prod_{\wt a=1}^{N_\gh}s_{b}(\vec\alpha\cdot\vec\phi+ c_{\wt a}\Lambda - {Q\over2i})^{\epsilon_{\wt a}} ,
\label{wtsb2sb}
}
where $s_{b}(\sigma)$ is the so-called double sine function defined by \cite{Kharchev:2001rs}
\be 
s_{b}(\sigma)
= \exp\[{1\over i} \int_0^\infty {dt \over t} \( {\sin{t\sigma} \over 2\sinh {bt \over 2} \sinh {b^{-1}t \over 2} } - {2\sigma\over t}\) \] .
\label{sb}
\ee
Note that we used \eqref{flavorvectorghost} in the last equality in \eqref{wtsb2sb}.
Plugging this expression into the above we find 
\beal{
Z^{\rm 1-loop}_\Lambda =& \prod_{\vec\alpha \in \Delta}\left\{s_{b}(\vec\alpha\cdot\vec\phi-{Q\over2i}) \prod_{\wt a=1}^{N_\gh}s_{b}(\vec\alpha\cdot\vec\phi+ c_{\wt a}\Lambda - {Q\over2i})^{\epsilon_{\wt a}}  \right\}.
}
The regularized one-loop determinant is obtained by sending $\Lambda$ to infinity
\beal{
[Z^{\rm 1-loop}]_{\rm reg}
=& \lim_{\Lambda\to \infty} Z^{\rm 1-loop}_{\Lambda}
= \prod_{\vec\alpha \in \Delta} \{[\omega(\vec\alpha\cdot\vec\phi-{Q\over2i})]_{\rm reg}\cdot s_{b}(\vec\alpha\cdot\vec\phi-{Q\over2i}) \},
}
where we set 
\be
\omega(\sigma):= \lim_{\Lambda\to \infty} \( \prod_{\wt a=1}^{N_\gh}s_{b}(\sigma+ c_{\wt a}\Lambda)^{\epsilon_{\wt a}} \).
\ee
The factor $\omega(\sigma)$, which arises due to the regularization, is a finite phase under the relation \eqref{flavorvectorghost}.
To show this we use the asymptotic expansion of the double sine function for a large $\sigma \sim \pm\infty$ \cite{Kharchev:2001rs}:
\be 
s_b(\sigma) = e^{\sgn(\sigma) {\pi i \over 2}(\sigma^2 + {b^2 + b^{-2} \over 12})   }(1 + \cO(e^{-a|\sigma|})), 
\label{sbxinfty}
\ee
where $a$ is a positive constant.
Employing this formula we can rewrite the factor as 
\be 
\omega(\sigma)= \lim_{\Lambda\to \infty} \prod_{\wt a=1}^{N_\gh}[ e^{{\pi i \over 2} \sgn(c_{\wt a})\epsilon_{\wt a} \{ (\sigma+ c_{\wt a}\Lambda)^2 + {b^2 + b^{-2} \over 12}   \}}] . 
\ee
This expression still contains the cut-off parameter $\Lambda$, but we can show that it can be removed by employing the relation \eqref{flavorvectorghost}:
\beal{
\omega(\sigma)
=&e^{{\pi i \over 2}\sum_{\wt a=1}^{N_\gh}\sgn(c_{\wt a})\epsilon_{\wt a} [\sigma^2 + {b^2 + b^{-2} \over 12} ] }. 
}
Further using the relation \eqref{flavorvectorghost2} we can write $\omega(\sigma)$ as 
\be
\omega(\sigma)
=e^{-{\pi i \over 2} \sgn(k_0) [ \sigma^2 + {b^2 + b^{-2} \over 12} ] }.
\label{omegaReg}
\ee

As a result we obtain 
\beal{
Z[\mbf S_b^3, G_{k_0}] 
=&{1 \over |W|} \int_{\mbf R^r} d^r\vec\phi\, e^{-\pi i k_0 |\vec\phi|^2} \prod_{\vec\alpha \in \Delta} \{ [\omega(\vec\alpha\cdot\vec\phi-{Q\over2i})]_{\rm reg}\cdot s_{b}(\vec\alpha\cdot\vec\phi-{Q\over2i}) \}.
}
Let us see the Chern-Simons level shift from the background dependent phase factor as shown by using the index theorem \cite{Witten:1988hf}.
In the current case, the product of the phase factors is computed as 
\beal{
\prod_{\vec\alpha \in \Delta} \omega(\vec\alpha\cdot\vec\phi-{Q\over2i})
=& e^{-{\pi i \over 2} \sgn(k_0) \sum_{\vec\alpha \in \Delta} [ (\vec\alpha\cdot\vec\phi-{Q\over2i})^2 + {b^2 + b^{-2} \over 12} ] } \nn
=& e^{-{\pi i \over 2} \sgn(k_0) [2h^\vee |\vec\phi|^2 -(\dim \mf g-r) (\half + {b^2 + b^{-2} \over 6}) ] },
}
where we used a formula known for a simple Lie algebra 
\be 
|\vec X|^2 
={1\over 2h^\vee}\sum_{\vec\alpha\in\Delta}(\vec\alpha\cdot\vec X)^2,
\label{completeSet}
\ee
with $\vec X$ an arbitrary vector in the $r$-dimensional Euclidean space and $h^\vee$ the dual Coxeter number. 
On the other hand, the product of the double sine functions simplifies to 
\beal{
\prod_{\vec\alpha \in \Delta} [s_{b}(\vec\alpha\cdot\vec\phi-{Q\over2i}) ]
=&\prod_{\vec\alpha \in \Delta_+} [s_{b}(\vec\alpha\cdot\vec\phi-{Q\over2i}) s_{b}(-\vec\alpha\cdot\vec\phi-{Q\over2i}) ] \nn
=&\prod_{\vec\alpha \in \Delta_+} [ 2 \sinh (\pi b \vec\alpha\cdot\vec\phi ) 2 \sinh (\pi b^{-1} \vec\alpha\cdot\vec\phi )],
}
where $\Delta_+$ is the set of the positive roots. 
Finally we obtain 
\beal{
Z[\mbf S_b^3, G_{k_0}] 
=&{ [e^{i \theta}]_{\rm reg} \over |W|} \int_{\mbf R^r} d^r\vec\phi\, e^{-\pi i k |\vec\phi|^2} \prod_{\vec\alpha \in \Delta_+} [ 2 \sinh (\pi b \vec\alpha\cdot\vec\phi ) 2 \sinh (\pi b^{-1} \vec\alpha\cdot\vec\phi )],
}
where $\theta={\pi \over 2} \sgn(k_0) (\dim \mf g-r) (\half + {b^2 + b^{-2} \over 6}) $, $k=k_0+ \sgn(k_0) h^\vee$. The level shift was confirmed. 

We have obtained a finite expression for the pure Chern-Simons partition function from the path-integral \eqref{defCSpf} employing the supersymmetric localization with supersymmetric regularization.
We still have degrees of freedom to add finite counter terms into the theory, which can be used for the total phase factor $[e^{i \theta}]_{\rm reg} $ to be rewritten as topological invariant. 
The prescription given in \cite{Witten:1988hf} is to add a gravitational Chern-Simons term, so we compute it for the squashed three spheres in appendix~\ref{gcs}. We expect that such a counter term cancels the dependence on a squashing parameter $b$, but as seen from results in appendix~\ref{gcs}, only a gravitational Chern-Simons term is not sufficient to remove the dependence on the squashing parameter.
This might be reasonable since it breaks supersymmetry. 
To restore the topological property, it would be important to add the supersymmetrized gravitational Chern-Simons term in the presented method. We leave it as a future work.

\subsection{For squashed lens space} 

We move on to discussing the case for a squashed lens space.
Since the $\mbb Z_{n}$ quotient is taken for the direction to which the anti-commutator of supercharges generates shift, the quotient does not break supersymmetry. Therefore the supersymmetric localization and regularization can be applied as in the case of a squashed three-sphere. 
Taking the $\mbb Z_{n}$ quotient creates a discrete 1-cycle, which becomes a non-trivial element of the fundamental group: $\pi_1(L(n,1))=\mbb Z_{n}$.
This generates new vacuum configurations of the system with holonomy turned on the discrete cycle $\gamma$  
\be 
{h\over n} = {1\over 2\pi} \oint_\gamma A,  
\ee
where $h$ is valued in the Cartan subalgebra of the gauge group. 
The components $h_s = \aver{h,H_s}$, where $H_s$ is a Cartan generator and $\aver{\cdot,\cdot}$ is the Killing form\footnote{The normalization of the Killing form in this paper is standard: $\aver{H_s,H_t}=\delta_{st}$ for the Cartan-Weyl basis.}, are discretized so that $h_s\in\mbb Z_n$ or $h_s=0,1,\cdots, n-1$ since $n\gamma$ becomes trivial in the lens space.  

The partition function on the squashed lens space is obtained by summing all over the possible saddle points with taking into account the holonomy as 
\beal{
Z[L_b(n,1), G_{k_0}] 
=&{1 \over n^r} \sum_{\vec h\in (\mbb Z_n)^r} {1 \over |W|}\int_{\mbf R^r} d^r\vec\phi\, e^{-S_0 } \[ Z^{\text{1-loop}}  \]_{\rm reg} .
}

We discuss the classical contribution from the Chern-Simons term later. 
The 1-loop determinant is computed by performing the Gaussian integration for truncated modes by the discrete action.  
The formal expression is given by \cite{Gang:2009qdj,Benini:2011nc,Imamura:2011wg} 
\beal{
Z^{\text{1-loop}}=& \prod_{\vec\alpha \in \Delta} \wt s_b^{(n,\vec\alpha\cdot\vec h)}(\vec\alpha\cdot\vec\phi-{b+ b^{-1}\over2i}),
}
where  
\beal{
\wt s_b^{(n,h)}(\sigma) 
=& \prod_{p,q \in \mbb Z_{\geq0} \atop  p-q\equiv h \!\!\mod n} {b(p+\half) + b^{-1} (q+\half)  -i\sigma \over b(p+\half) + b^{-1} (q+\half) + i\sigma }. 
\label{wtsbnh}
}
This 1-loop determinant contains ultra-violet divergence, and we regularize it by adding the same regulator chiral multiplets as used in the case of the three sphere.
Redoing the calculation we obtain the 1-loop determinant as  
\beal{
Z^{\text{1-loop}}_{\Lambda} 
=& \prod_{\vec\alpha \in \Delta}\left\{\wt s^{(n,\vec\alpha\cdot\vec h)}_{b}(\vec\alpha\cdot\vec\phi-{Q\over2i}) \prod_{\wt a=1}^{N_\gh} \( \wt s^{(n,\vec\alpha\cdot\vec h)}_{b}(\vec\alpha\cdot\vec\phi+ c_{\wt a}\Lambda - {Q\over2i} ) \)^{\epsilon_{\wt a}} \right\} . 
}
This must give a finite result. 
To see this we first rewrite $\wt s_b^{(n,h)}(x)$ in terms of $\wt s_b(x)$ as 
\be 
\wt s_b^{(n,h)}(x) 
= \prod_{l=0}^{n-1} \wt s_b({x \over n} + ib\aver{l+h }_n + ib^{-1}\aver{ l }_n ),
\ee
where $\aver{m}_n:= {1\over n} ([m]_n + \half ) -\half$ with $[m]_n$ the remainder of $m$ divided by $n$. 
Then as seen in the previous subsection, we can replace the infinite product $\wt s_b$ with a finite function $s_b$. 
Therefore 
\beal{
Z^{\text{1-loop}}_{\Lambda} 
=& \prod_{\vec\alpha \in \Delta}\left\{s^{(n,\vec\alpha\cdot\vec h)}_{b}(\vec\alpha\cdot\vec\phi-{Q\over2i})\prod_{\wt a=1}^{N_\gh} \bigg(s^{(n,\vec\alpha\cdot\vec h)}_{b}(\vec\alpha\cdot\vec\phi+ c_{\wt a}\Lambda - {Q\over2i} ) \bigg)^{\epsilon_{\wt a}}  \right\},
}
where 
\beal{
s_b^{(n,h)}(\sigma)=& \prod_{l=0}^{n-1} s_b({\sigma \over n} + ib\aver{k+h }_n + ib^{-1}\aver{ k }_n )   .
}
Then the regularized 1-loop determinant $[Z^{\rm 1-loop}]_{\rm reg}$ is obtained by taking the limit $\Lambda\to\infty$ for $Z^{\text{1-loop}}_{\Lambda} $. 
\beal{
[Z^{\rm 1-loop}]_{\rm reg}
=& \prod_{\vec\alpha \in \Delta}\bigg\{[ \omega^{(n,\vec\alpha\cdot\vec h)}(\vec\alpha\cdot\vec\phi-{Q\over2i})]_{\rm reg} \cdot s^{(n,\vec\alpha\cdot\vec h)}(\vec\alpha\cdot\vec\phi-{Q\over2i})\bigg\},
}
where 
\beal{ 
\omega^{(n,h)}(\sigma):=& \lim_{\Lambda\to \infty} \prod_{\wt a=1}^{N_\gh} \bigg(s^{(n,\vec\alpha\cdot\vec h)}_{b}(\sigma+ c_{\wt a}\Lambda ) \bigg)^{\epsilon_{\wt a}} .
}
Let us compute the phase factor. 
Using \eqref{sbxinfty} we find  
\beal{
\omega^{(n,h)}(\sigma)
=&\lim_{\Lambda\to \infty} \prod_{l=0}^{n-1} \prod_{\wt a=1}^{N_\gh}  e^{{\pi i \over 2}\epsilon_{\wt a}\sgn(c_{\wt a})(({\sigma + c_{\wt a}\Lambda \over n} + ib\aver{k+h }_n + ib^{-1}\aver{ k }_n)^2 +{b^2 + b^{-2} \over 12})   }. 
}
This expression still contains $\Lambda$, which disappears by using relations \eqref{flavorvectorghost}. 
\beal{
\omega^{(n,h)}(\sigma)
=&e^{ {\pi i\over 2n}\sum_{\wt a=1}^{N_\gh}\sgn(c_{\wt a})\epsilon_{\wt a} \{ \sigma ^2 - ([h]_n([h]_n-n) + \frac{n^2-1}{6 }) - \frac{1}{12 } ( b^2+ b^{-2}) \} },
}
where we computed by using 
\beal{
\sum_{l=0}^{n-1}\aver{ l }_n= 0, \quad 
\sum_{l=0}^{n-1}\aver{ l }_n^2= \frac{n^2-1}{12 n}, \quad 
\sum_{l=0}^{n-1}\aver{ l }_n\aver{ l+h }_n= \frac{6 [h]_n^2-6 [h]_n n+n^2-1}{12 n}.  
}
Further using the relation \eqref{flavorvectorghost2} we obtain 
\beal{
\omega^{(n,h)}(\sigma)
=&e^{ -{\pi i\over 2n}\sgn(k_0) \{ \sigma ^2 - [h]_n([h]_n-n) - \frac{n^2-1}{6 } - \frac{1}{12 } ( b^2+ b^{-2}) \} }.
\label{omegaRegLens}
}
In the computation prescribed in \cite{Witten:1988hf} this phase factor corresponds to the $\eta$ invariant. The $\eta$-invariant for an unsquashed lens space under holonomy background is given by \cite{CisnerosMolina2001TheO}
\beal{
\eta(L_1(n,1);h)
=&\frac{6 h^2-6 h n+n^2-1}{6 n}. 
}
We also confirm this result from a direct computation in appendix \ref{eta-invariant} . Thus the relation of the phase factor computed in the supersymmetric localization and the $\eta$-invariant is 
\be 
\omega^{(n,h)}(\sigma)/\omega^{(1,0)}(\sigma)
=e^{{\pi i\over 2}\sgn(k_0)\eta(L_1(n,1);[h]_n) } . 
\ee

As a result we obtain   
\beal{
Z[L_b(n,1), G_{k_0}] 
=&{1\over n^r } \sum_{\vec h\in (\mbb Z_n)^r}{1 \over |W|} \int_{\mbf R^r} d^r\vec\phi\, e^{-S_0 }  \prod_{\vec\alpha \in \Delta}\bigg\{ [\omega^{(n,\vec\alpha\cdot\vec h)}(\vec\alpha\cdot\vec\phi-{Q\over2i})]_{\rm reg} \cdot s_b^{(n,\vec\alpha\cdot\vec h)}(\vec\alpha\cdot\vec\phi-{Q\over2i})\bigg\} .
\label{middleZ}
}
Let us compute the product of the background dependent phase factors.
\beal{
\prod_{\vec\alpha \in \Delta}\omega^{(n,\vec\alpha\cdot\vec h)}(\vec\alpha\cdot\vec\phi-{Q\over2i})
=& e^{-{\pi i \over 2n } \sgn(k_0) \sum_{\vec\alpha \in \Delta} [ (\vec\alpha\cdot\vec\phi-{Q\over2i})^2  - [\vec\alpha\cdot\vec h]_n([\vec\alpha\cdot\vec h]_n-n) - \frac{n^2-1}{6 } - \frac{1}{12 } ( b^2+ b^{-2})  ] } \nn
=& e^{-{\pi i \over 2n} \sgn(k_0) [2h^\vee |\vec\phi|^2 + (\dim \mf g-r) (-\frac{n^2-1}{6 }-\half - {b^2 + b^{-2} \over 6}) -\sum_{\vec\alpha \in \Delta} [\vec\alpha\cdot\vec h]_n([\vec\alpha\cdot\vec h]_n-n) ] }
\notag ,}
where we computed the holonomy independent part in the same way as in section \ref{squashedS3} using \eqref{completeSet}. 
We compute the holonomy dependent part in the following manner. 
\beal{
\sum_{\vec\alpha \in \Delta} [\vec\alpha\cdot\vec h]_n([\vec\alpha\cdot\vec h]_n-n) 
=&\sum_{\vec\alpha \in \Delta_+}\{ [\vec\alpha\cdot\vec h]_n([\vec\alpha\cdot\vec h]_n-n) +[-\vec\alpha\cdot\vec h]_n([-\vec\alpha\cdot\vec h]_n-n) \} \nn
=&2\sum_{\vec\alpha \in \Delta_+}[\vec\alpha\cdot\vec h]_n([\vec\alpha\cdot\vec h]_n-n),
}
where we used $[m]_n+[-m]_n = n(1-\delta_{m0})$. 
Therefore 
\beal{
\exp\[{\pi i \over 2n} \sgn(k_0) \sum_{\vec\alpha \in \Delta} [\vec\alpha\cdot\vec h]_n([\vec\alpha\cdot\vec h]_n-n) ]\]
=& \exp\[{\pi i \over n} \sgn(k_0) \sum_{\vec\alpha \in \Delta_+} [\vec\alpha\cdot\vec h]_n([\vec\alpha\cdot\vec h]_n-n) ]\] \nn
=& \exp\[{\pi i \over n} \sgn(k_0) \sum_{\vec\alpha \in \Delta_+} ([\vec\alpha\cdot\vec h]_n)^2(1-n) ]\], \notag
}
since $(-1)^m= (-1)^{m^2}$ for an integer $m$.
Denote $(x)_n$ by the quotient of $x$ divided by $n$. Then 
\beal{
\exp\[{\pi i \over n} \sgn(k_0) \sum_{\vec\alpha \in \Delta_+} ([\vec\alpha\cdot\vec h]_n)^2(1-n) ]\]
=&\exp\[{\pi i \over n} \sgn(k_0) \sum_{\vec\alpha \in \Delta_+} (\vec\alpha\cdot\vec h - n (\vec\alpha\cdot\vec h)_n)^2(1-n) \] \nn
=&\exp\[{\pi i \over n} \sgn(k_0) \sum_{\vec\alpha \in \Delta_+} (\vec\alpha\cdot\vec h)^2(1-n) \] \nn
=&\exp\[{\pi i \over n} \sgn(k_0) h^\vee |\vec h|^2(1-n) \] ,
}
where in the last we used \eqref{completeSet}. 
Therefore 
\beal{
\prod_{\vec\alpha \in \Delta}[\omega^{(n,\vec\alpha\cdot\vec h)}(\vec\alpha\cdot\vec\phi-{Q\over2i})]_{\rm reg}
=&[e^{i\theta^{(n)} }]_{\rm reg}  e^{-{\pi i \over n} \sgn(k_0) h^\vee ( |\vec\phi|^2 +(n-1)|\vec h|^2 ) },
\notag }
where $\theta^{(n)}={\pi \over 2n} \sgn(k_0) (\dim \mf g-r) (\frac{n^2-1}{6 }+\half + {b^2 + b^{-2} \over 6}) $.

On the other hand, the product of the double sine functions can be computed as  
\beal{
\prod_{\vec\alpha \in \Delta} [s^{(n,\vec\alpha\cdot\vec h)}_{b}(\vec\alpha\cdot\vec\phi-{Q\over2i}) ]
=&\prod_{\vec\alpha \in \Delta_+} [s^{(n,\vec\alpha\cdot\vec h)}_{b}(\vec\alpha\cdot\vec\phi-{Q\over2i}) s^{(n,-\vec\alpha\cdot\vec h)}_{b}(-\vec\alpha\cdot\vec\phi-{Q\over2i}) ] \nn
=& \prod_{\vec\alpha\in\Delta_+}[ 2 \sinh(\pi{b^{-1}\vec\alpha\cdot\vec\phi + i \vec\alpha\cdot\vec h \over n} ) 2\sinh (\pi{b\vec\alpha\cdot\vec\phi -i \vec\alpha\cdot\vec h  \over n})].
}

Finally we obtain 
\beal{
Z[L_b(n,1), G_{k_0}] 
=& {[e^{i\theta^{(n)} }]_{\rm reg}  \over n^r } \sum_{\vec h\in (\mbb Z_n)^r}{1 \over |W|} \int_{\mbf R^r} d^r\vec\phi\, e^{-S_0-{\pi i \over n} \sgn(k_0) h^\vee ( |\vec\phi|^2 +(n-1)|\vec h|^2 ) } \nn
& \times \prod_{\vec\alpha \in \Delta}\bigg\{2 \sinh(\pi{b^{-1}\vec\alpha\cdot\vec\phi + i \vec\alpha\cdot\vec h \over n} ) 2\sinh (\pi{b\vec\alpha\cdot\vec\phi -i \vec\alpha\cdot\vec h  \over n}) \bigg\}. 
\label{pfGk}
}
From this result we can induce the classical contribution of the Chern-Simons term on the lens space, which must give the correct level shift as seen in the case of the squashed three sphere. 
Such a contribution is uniquely determined as 
\be 
S_0 =  {\pi i k_0 \over n} \{|\vec\phi|^2 +(n-1)|\vec h|^2 \} .
\label{classicalPartLens}
\ee
This result is consistent with the one given in \cite{Imamura:2012rq,Imamura:2013qxa} rather than in \cite{2002math......9403H,Brini:2008ik,Gang:2009qdj,Alday:2013lba}. 

Although our derivation assumes that a Lie group for gauge symmetry is simple, it is not difficult to extend the result to a semi-simple Lie group and the one including abelian factors. 
In the following section we study the level rank duality with respect to the unitary and special unitary groups, so let us derive the lens space partition function for the $U(N)^{(0)}_{k_0,k_0'}$ pure Chern-Simons theory whose action is 
\be 
\int_{L_b(n,1)}\[ {i k_0\over 4\pi}\tr( A \wedge dA + {2 \over 3} A\wedge A \wedge A) + {i k_0'\over 4\pi}\tr A  \wedge d \tr A \]. 
\ee
The partition function corresponding to \eqref{middleZ} is written as 
\beal{
Z[L_b(n,1), U(N)^{(0)}_{k_0,k_0'}] 
=&{1\over n^N } \hsum{1 \over N!} \int_{\mbf R^N} d^N\vec\phi\, e^{-{\pi i k_0 \over n} \sum_{s=1}^N\{\phi_s^2 +(n-1) h_s^2 \}-{\pi i k_0' \over n} \{\phi_D^2 +(n-1) h_D^2 \} } \nn
& \times \prod_{1\leq s,t \leq N \atop s\not=t}\bigg\{ [\omega^{(n,h_{st})}(\phi_{st}-{Q\over2i})]_{\rm reg} \cdot s_b^{(n,h_{st})}(\phi_{st}-{Q\over2i})\bigg\} ,
}
where we used notation 
\be 
X_D=\sum_{s=1}^N X_s,~~ X_{st}=X_s - X_t.
\label{DiagDef}
\ee
The products of the background dependent phase factors is computed as 
\beal{
\prod_{1\leq s,t \leq N \atop s\not=t}[\omega^{(n,h_{st})}(\phi_{st}-{Q\over2i})]_{\rm reg}
=&[e^{i\theta^{(n)} }]_{\rm reg}  e^{-{\pi i \over n} \sgn(k_0) \[ N\sum_{s=1}^N\{\phi_s^2 +(n-1) h_s^2 \} - \{\phi_D^2 +(n-1) h_D^2 \} \] },
\notag }
where $\theta^{(n)}={\pi \over 2n} \sgn(k_0) (N^2 -N) (\frac{n^2-1}{6 }+\half + {b^2 + b^{-2} \over 6}) $.
Here we used the following formula 
\be 
N\sum_{s=1}^N X_s^2 = \sum_{1\leq s,t \leq N \atop s<t} (X_{st})^2 + X_D^2 ,
\ee
instead of \eqref{completeSet}. 
Finally we obtain 
\beal{
Z[L_b(n,1), U(N)^{(0)}_{k_0,k_0'}] 
=& {[e^{i\theta^{(n)} }]_{\rm reg} \over n^N } \hsum{1 \over N!} \int_{\mbf R^N} d^N\vec\phi\, e^{-{\pi i k \over n} \sum_{s=1}^N\{\phi_s^2 +(n-1) h_s^2 \}-{\pi i k' \over n} \{\phi_D^2 +(n-1) h_D^2 \} } \nn
& \times \prod_{1\leq s,t \leq N \atop s<t}\bigg\{2 \sinh(\pi{b^{-1}\phi_{st} + i h_{st} \over n} ) 2\sinh (\pi{b\phi_{st} -i h_{st}  \over n}) \bigg\}, 
\label{pfUNkk'}
}
where $k=k_0+ \sgn(k_0)N, ~ k'=k'_0 -\sgn(k_0)$. 

We still have degrees of freedom to add a finite counter term to make the overall phase $[e^{i\theta^{(n)} }]_{\rm reg}$ topologically invariant. 
We expect that if this were done then an arbitrary level rank duality could be tested by using this partition function including the overall phase.\footnote{The partition functions \eqref{pfGk} and \eqref{pfUNkk'} are sufficient to test a duality only for the absolute value. }
Since we do not determine such a correct counter term in this paper, we give up calculating the topological phase from the first principle.
Instead in the following section we show that by replacing $[e^{i\theta^{(n)} }]_{\rm reg}$ with a certain phase $e^{i\Theta(N,k,k') }$ the well-known level rank duality as well as the newly proposed one in \cite{Hsin:2016blu} are confirmed including the total phase by numerical calculation with high precision. 
Note that the phase factor $e^{i\Theta(N,k,k') }$ cannot be obtained just by using the corresponding WZNW model and its known modular matrices. 
This finding is another new achievement in this paper, and discussing it with the level rank duality is the topic in the next section. 
We use notation $U(N)_{k,k'}=U(N)^{(0)}_{k_0,k_0'}$ from the next section.

\section{Matrix models\label{sec:MatrixModel}}
As we have explained in the previous sections, our aim in this paper is to find the correct phase for various level-rank dualities. In this section, we study the matrix model. The result will be used for studying the level-rank dualities and the phase in section \ref{sec:Duality}.

In the rest of this paper, we focus on the case when the gauge group is ${\rm U}\left(N\right)_{k,k'}$ or ${\rm SU}\left(N\right)_{k}$. The matrix model for ${\rm U}\left(N\right)_{k,k'}$ gauge group is \eqref{pfUNkk'}. We define the matrix model without the phase $[e^{i\theta^{(n)}}]_{{\rm reg}}$
\begin{align}
\tilde{Z}\left[L_{b}\left(n,1\right),{\rm U}\left(N\right)_{k,k'}\right]= & \frac{1}{n^{N}}\hsum\int\frac{d^{N}\phi}{N!}e^{-\sum_{s}^{N}\frac{i\pi k}{n}\left[\phi_{s}^{2}+\left(n-1\right)h_{s}^{2}\right]}e^{-\frac{i\pi k'}{n}\left[\phi_{{\rm D}}^{2}+\left(n-1\right)h_{{\rm D}}^{2}\right]}\nonumber \\
 & \times\prod_{s<t}^{N}2\sinh\frac{\pi}{n}\left(b^{-1}\phi_{st}+ih_{st}\right)2\sinh\frac{\pi}{n}\left(b\phi_{st}-ih_{st}\right),\label{eq:UCSPFDef}
\end{align}
where $\phi_{{\rm D}}$ and $h_{{\rm D}}$ are defined in \eqref{DiagDef}. Similarly, the matrix model \eqref{pfGk} for ${\rm SU}\left(N\right)_{k}$ gauge group without the phase $[e^{i\theta^{(n)}}]_{{\rm reg}}$ is
\begin{align}
\tilde{Z}\left[L_{b}\left(n,1\right),{\rm SU}\left(N\right)_{k}\right]= & \frac{1}{n^{N-1}}\hsum\delta_{h_{{\rm D}},0}^{{\rm mod}\thinspace n}\int\frac{d^{N}\phi}{N!}\delta\left(\phi_{{\rm D}}\right)e^{-\sum_{s}^{N}\frac{i\pi k}{n}\left[\phi_{s}^{2}+\left(n-1\right)h_{s}^{2}\right]}\nonumber \\
 & \times\prod_{s<t}^{N}2\sinh\frac{\pi}{n}\left(b^{-1}\phi_{st}+ih_{st}\right)2\sinh\frac{\pi}{n}\left(b\phi_{st}-ih_{st}\right),\label{eq:SUCSPFdef}
\end{align}
where
\begin{equation}
\delta_{a,b}^{{\rm mod}\thinspace n}=\sum_{m\in\mathbb{Z}}\delta_{a,b+nm}.\label{eq:ModDeltaDef}
\end{equation}
We do not write $L_{b}\left(n,1\right)$ explicitly because these parameters are always equal for dual theories in this paper.

One of the main result is that
\begin{align}
Z\left[{\rm U}\left(N\right)_{k,k'}\right] & =e^{i\Theta\left(N,k\right)+i\Theta\left(1,k+k'N\right)-i\Theta\left(1,k\right)}\tilde{Z}\left[{\rm U}\left(N\right)_{k,k'}\right],\nonumber \\
Z\left[{\rm SU}\left(N\right)_{k}\right] & =e^{i\Theta\left(N,k\right)-i\Theta\left(1,k\right)}\tilde{Z}\left[{\rm SU}\left(N\right)_{k}\right],\label{eq:PhasePFdef}
\end{align}
where
\begin{equation}
\Theta\left(N,k\right)={\rm sgn}\left(k\right)\frac{\pi nN^{2}}{4}+\frac{\pi\left(b^{2}+b^{-2}\right)}{12nk}\left(N^{3}-N\right)-\frac{\pi\left(n-n^{-1}\right)}{6k}\left(N^{3}-N\right),\label{eq:AllPhaseDef}
\end{equation}
makes dualities complete. The derivation of this result is in section \ref{sec:Duality}.%
\footnote{ 
If we naively compute the corresponding value from the known modular matrices for the 2d SU($N$) WZNW model by $\ms Z\left[{\rm SU}\left(N\right)_{k}\right]= (ST^nS)_{00}$ as done in \cite{2002math......9403H}, for example, the result is   
$\ms Z\left[{\rm SU}\left(N\right)_{k}\right]=e^{ -({n\pi i \over N} +{4\pi i \over n k}){N^3 - N \over 12}} \tilde{\ms Z}\left[{\rm SU}\left(N\right)_{k}\right]$, where 
\beal{
\tilde{\ms Z}\left[{\rm SU}\left(N\right)_{k}\right]=& \frac{1}{n^{N-1}}\hsum\delta_{h_{{\rm D}},0}^{{\rm mod}\thinspace n}\int\frac{d^{N}\phi}{N!}\delta\left(\phi_{{\rm D}}\right)e^{-\sum_{s}^{N}\frac{i\pi k}{n}\left[\phi_{s}^{2}-h_{s}^{2}\right]}\nonumber \\
 & \times \prod_{s<t}^{N}2\sinh\frac{\pi}{n}\left(\phi_{st}+ih_{st}\right)2\sinh\frac{\pi}{n}\left(\phi_{st}-ih_{st}\right). \notag
}
Thus both the overall phase and the relative phase depending on background holonomy are different from our result. 
We will comment on this point in the discussion section.
}

The matrix model satisfies
\begin{align}
\tilde{Z}\left[{\rm U}\left(N\right)_{k,k'}\right] & =\left(\tilde{Z}\left[{\rm U}\left(N\right)_{-k,-k'}\right]\right)^{*},\nonumber \\
\tilde{Z}\left[{\rm SU}\left(N\right)_{k,}\right] & =\left(\tilde{Z}\left[{\rm SU}\left(N\right)_{-k,}\right]\right)^{*}.\label{eq:knMinus}
\end{align}
Here we used the invariance of the matrix model under $b\rightarrow b^{-1}$. This invariance comes from the fact that $L_{b}\left(n,1\right)$ and $L_{b^{-1}}\left(n,1\right)$ are the same manifolds. The invariance can also be explicitly seen after applying the Fermi gas formalism in section \ref{subsec:Numerical}.

In the rest of this section, we study some aspects of the matrix models. In section \ref{subsec:UtoSU}, we study the relation of the matrix models between ${\rm U}\left(N\right)_{k,k'}$ and ${\rm SU}\left(N\right)_{k}$. In section \ref{subsec:Numerical}, we apply the Fermi gas formalism to the matrix model. The results will be used for studying level-rank dualities in section \ref{sec:Duality}.

\subsection{${\rm U}\left(N\right)_{k,k'}$ and ${\rm SU}\left(N\right)_{k}$\label{subsec:UtoSU}}

The Lagrangian of ${\rm U}\left(N\right)_{k,k'}$ gauge field decomposes as \cite{Hsin:2016blu}\footnote{Since ${\rm U}\left(1\right)_{k,k'}={\rm U}\left(1\right)_{k+k',0}$, we set $k'=0$ for the abelian case.}
\begin{equation}
{\rm U}\left(N\right)_{k,k'}=\frac{{\rm U}\left(1\right)_{kN+k'N^{2}}\times{\rm SU}\left(N\right)_{k}}{\mathbb{Z}_{N}}.\label{eq:UtoSUgroup}
\end{equation}
This relation can be seen in terms of the matrix model. In this section, we show the relation\footnote{The first factor $N$ in the right hand side corresponds to $\mathbb{Z}_{N}$ in the right hand side of \eqref{eq:UtoSUgroup} because the matrix model should be divided by the number of elements of the Weyl group.}
\begin{equation}
\tilde{Z}\left[{\rm U}\left(N\right)_{k,k'}\right]=N\times\tilde{Z}\left[{\rm U}\left(1\right)_{kN+k'N^{2}}\right]\tilde{Z}\left[{\rm SU}\left(N\right)_{k}\right],\label{eq:UtoSU}
\end{equation}
when $\gcd\left(N,n\right)=1$. By using this relation, we can also relate the ${\rm U}\left(N\right)_{k,k'}$ matrix model with and without the diagonal part
\begin{equation}
\tilde{Z}\left[{\rm U}\left(N\right)_{k,k'}\right]=\frac{\tilde{Z}\left[{\rm U}\left(1\right)_{kN+k'N^{2}}\right]}{\tilde{Z}\left[{\rm U}\left(1\right)_{kN}\right]}\tilde{Z}\left[{\rm U}\left(N\right)_{k,0}\right],\label{eq:w-woDiag}
\end{equation}
when $\tilde{Z}\left[{\rm U}\left(1\right)_{kN}\right]\neq0$.

To obtain \eqref{eq:UtoSU}, we need to decompose both the integral over the Cartan of the gauge group and the sum over $h_{i}$ labeling the flat gauge fields. The former integral can be decomposed for any cases, while the latter sum can be decomposed only when $\gcd\left(N,n\right)=1$.

We start with the matrix model \eqref{eq:UCSPFDef} for ${\rm U}\left(N\right)_{k,k'}$. The integral over the Cartan becomes
\begin{align}
 & \int\frac{d^{N}\phi}{N!}e^{-\sum_{s}^{N}\frac{i\pi k}{n}\phi_{s}^{2}}e^{-\frac{i\pi k'}{n}\phi_{{\rm D}}^{2}}\prod_{s<t}^{N}2\sinh\frac{\pi}{n}\left(b^{-1}\phi_{st}+ih_{st}\right)2\sinh\frac{\pi}{n}\left(b\phi_{st}-ih_{st}\right)\nonumber \\
 & =N\int d\tilde{\phi}e^{-\frac{i\pi\left(kN+k'N^{2}\right)}{n}\tilde{\phi}^{2}}\int\frac{d^{N}\phi}{N!}\delta\left(\phi_{{\rm D}}\right)e^{-\sum_{s}^{N}\frac{i\pi k}{n}\phi_{s}^{2}} \nonumber \\
& \quad \times \prod_{s<t}^{N}2\sinh\frac{\pi}{n}\left(b^{-1}\phi_{st}+ih_{st}\right)2\sinh\frac{\pi}{n}\left(b\phi_{st}-ih_{st}\right).\label{eq:Decom1}
\end{align}
Next, we consider the decomposition of the sum over $h_{i}$. Because
\begin{equation}
e^{-\sum_{s}^{N}\frac{i\pi k}{n}\left(n-1\right)h_{s}^{2}}e^{-\frac{i\pi k'}{n}\left(n-1\right)h_{{\rm D}}^{2}}\prod_{s<t}^{N}2\sinh\frac{\pi}{n}\left(b^{-1}\phi_{st}+ih_{st}\right)2\sinh\frac{\pi}{n}\left(b\phi_{st}-ih_{st}\right),
\end{equation}
is invariant under the shift
\begin{equation}
\left\{ h_{i}\right\} _{1\leq i\leq N}\rightarrow\left\{ h_{i}+n\right\} _{1\leq i\leq N},
\end{equation}
we can use \eqref{eq:DecomLemma} when $\gcd\left(N,n\right)=1$.
The matrix model becomes
\begin{align}
\tilde{Z}\left[{\rm U}\left(N\right)_{k,k'}\right]= & N\int d\tilde{\phi}e^{-\frac{i\pi\left(kN+k'N^{2}\right)}{n}\tilde{\phi}^{2}}\frac{1}{n^{N}}\sum_{\tilde{h}=0}^{n-1}\sum_{\left\{ h_{i}\right\} _{1\leq i\leq N}=0}^{n-1}\delta_{h_{{\rm D}},0}^{{\rm mod}\thinspace n}\nonumber \\
 & \times\int\frac{d^{N}\phi}{N!}\delta\left(\phi_{{\rm D}}\right)e^{-\sum_{s}^{N}\frac{i\pi k}{n}\left[\phi_{s}^{2}+\left(n-1\right)\left(h_{s}+\tilde{h}\right)^{2}\right]}e^{-\frac{i\pi k'}{n}\left(n-1\right)\left(h_{{\rm D}}+N\tilde{h}\right)^{2}}\nonumber \\
 & \times\prod_{s<t}^{N}2\sinh\frac{\pi}{n}\left(b^{-1}\phi_{st}+ih_{st}\right)2\sinh\frac{\pi}{n}\left(b\phi_{st}-ih_{st}\right).
\end{align}
After a short computation, we obtain \eqref{eq:UtoSU}.

\subsection{Fermi gas-like approach\label{subsec:Numerical}}

Though the matrix model is finite dimensional integral, it is still hard to study analytically. The novel technique called the Fermi gas formalism was first introduced for the ABJM theory on $S^{3}$ in \cite{Marino:2011eh} and extended for rank deformed cases in \cite{Matsumoto:2013nya,Honda:2013pea,Honda:2014npa}. Since the pure Chern-Simons theory can be considered to be the rank deformed ABJ(M) theory of which one rank is zero, we expect that we can also apply the Fermi gas formalism to the pure Chern-Simons theory. In this section, we apply the Fermi gas formalism to the pure Chern-Simons theory with ${\rm U}\left(N\right)_{k,k'}$ and ${\rm SU}\left(N\right)_{k}$ gauge group on lens space (see also \cite{Yokoyama:2016ktm}).

The Fermi gas formalism is useful from various aspects. 
For instance, the Fermi gas formalism can be used for studying the matrix model numerically. Here we use the word ``numerical'' in the sense that we can obtain the exact value of the matrix model for fixed parameters (using, for example, \textsl{Mathematica}). The reason we can obtain the exact value is that the Fermi gas formalism for pure Chern-Simons theory completely perform the integration of the matrix model. Therefore, what we need to calculate is just the sum over $h_{i}$.

To apply the Fermi gas formalism, it is crucial that the integrand of the matrix model does not include a factor which is a function of $\phi_{{\rm D}}$. However, the both $\tilde{Z}\left[{\rm U}\left(N\right)_{k,k'}\right]$ and $\tilde{Z}\left[{\rm SU}\left(N\right)_{k}\right]$ include such factor. Therefore, we need to eliminate these factors before applying the Fermi gas formalism. For this purpose, we introduce a function
\begin{align}
\mathcal{Z}_{k,N}^{\left[f\left(h_{{\rm D}}\right)\right]}= & \frac{1}{n^{N}}\hsum f\left(h_{{\rm D}}\right)\int\frac{d^{N}\phi}{N!}e^{-\sum_{s}^{N}\frac{i\pi k}{n}\left[\phi_{s}^{2}+\left(n-1\right)h_{s}^{2}\right]}\nonumber \\
 & \times\prod_{s<t}^{N}2\sinh\frac{\pi}{n}\left(b^{-1}\phi_{st}+ih_{st}\right)2\sinh\frac{\pi}{n}\left(b\phi_{st}-ih_{st}\right),\label{eq:MMgeneral}
\end{align}
and define
\begin{align}
\mathcal{Z}_{k,N}^{\delta} & =\left.\mathcal{Z}_{k,N}^{\left[f\left(h_{{\rm D}}\right)\right]}\right|_{f\left(h_{{\rm D}}\right)=\delta_{h_{{\rm D}},0}^{{\rm mod}\thinspace n}},\quad\mathcal{Z}_{k,N}^{k'}=\left.\mathcal{Z}_{k,N}^{\left[f\left(h_{{\rm D}}\right)\right]}\right|_{f\left(h_{{\rm D}}\right)=e^{-\frac{i\pi k'}{n}\left(n-1\right)h_{{\rm D}}^{2}}}.
\end{align}
Using \eqref{eq:Decom1}, they are related to the matrix models associated to ${\rm U}\left(N\right)_{k,k'}$ and ${\rm SU}\left(N\right)_{k}$ as
\begin{align}
\tilde{Z}\left[{\rm U}\left(N\right)_{k,k'}\right] & =i^{{\rm sgn}\left(k\right)\frac{1}{2}-{\rm sgn}\left(k+k'N\right)\frac{1}{2}}\sqrt{\frac{\left|k\right|}{\left|k+k'N\right|}}\mathcal{Z}_{k,N}^{k'},\label{eq:UFGF}
\end{align}
and
\begin{align}
\tilde{Z}\left[{\rm SU}\left(N\right)_{k}\right] & =i^{{\rm sgn}\left(k\right)\frac{1}{2}}\sqrt{\frac{n\left|k\right|}{N}}\mathcal{Z}_{k,N}^{\delta}.\label{eq:SUFGF}
\end{align}
Therefore, it is sufficient to apply the Fermi gas formalism to \eqref{eq:MMgeneral}. In the rest of this section, we carry out this task. The result is \eqref{eq:FGFres}. 

We assume $k\geq1$ for simplicity. The negative $k$ case can be computed in the similar way. First, by applying the determinant formula
\begin{equation}
\prod_{s<t}^{N}2\sinh\frac{\pi}{n}\left(\alpha_{s}-\alpha_{t}\right)=\det\left(e^{\frac{2\pi}{n}\sigma_{s}\alpha_{t}}\right)_{s,t},\label{eq:DetFormula}
\end{equation}
to the two $\sinh$ factors in \eqref{eq:MMgeneral}, we obtain
\begin{align}
\mathcal{Z}_{k,N}^{\left[f\left(h_{{\rm D}}\right)\right]} & =\frac{1}{k^{N}}\hsum\int\frac{d^{N}\phi}{N!}f\left(h_{{\rm D}}\right)e^{-\sum_{s}^{N}\frac{i\pi k}{n}\left[\phi_{s}^{2}+\left(n-1\right)h_{s}^{2}\right]}\nonumber \\
 & \quad\times\det\left(e^{\frac{2\pi}{n}\sigma_{s}\left(b^{-1}\phi_{t}+ih_{t}\right)}\right)_{s,t}\det\left(e^{\frac{2\pi}{n}\sigma_{t}\left(b\phi_{s}-ih_{s}\right)}\right)_{s,t},
\end{align}
where
\begin{equation}
\sigma_{s}=\frac{N+1}{2}-s.
\end{equation}

The calculation becomes easier by using the knowledge of the quantum mechanics \cite{Kiyoshige:2016lno,Moriyama:2017gye,Moriyama:2015asx,Moriyama:2016xin,Moriyama:2016kqi,Kubo:2020qed}. The notation is as follows. Commutation relation of position operator $\hat{x}$ and momentum operator $\hat{p}$ is $\left[\hat{x},\hat{p}\right]=i\hbar$, where $\hbar=\frac{n}{2\pi k}$. $\ket{\cdot}$ denotes a position eigenvector. We also introduce a symbol $\kket{\cdot}$ denoting a momentum eigenvector. The inner products of these vectors are
\begin{align}
 & \braket{x_{1}|x_{2}}=\delta\left(x_{1}-x_{2}\right),\quad\bbrakket{p_{1}|p_{2}}=\delta\left(p_{1}-p_{2}\right),\quad\brakket{x|p}=\frac{1}{\sqrt{2\pi\hbar}}e^{\frac{i}{\hbar}xp}.\label{eq:Normalization}
\end{align}

With this notation, the matrix model can be written as
\begin{align}
\mathcal{Z}_{k,N}^{\left[f\left(h_{{\rm D}}\right)\right]}= & \frac{1}{k^{N}}\hsum\int\frac{d^{N}\phi}{N!}f\left(h_{{\rm D}}\right)e^{-\sum_{s}^{N}\frac{i}{2\hbar}\left[\phi_{s}^{2}+\left(n-1\right)h_{s}^{2}\right]}\nonumber \\
 & \times\det\left(e^{\frac{2\pi i}{n}h_{t}\sigma_{s}}\bbraket{\frac{i}{bk}\sigma_{s}|\phi_{t}}\right)_{s,t}\det\left(e^{-\frac{2\pi i}{n}h_{s}\sigma_{t}}\brakket{\phi_{s}|-\frac{ib}{k}\sigma_{t}}\right)_{s,t},
\end{align}
After putting the Fresnel factor into the second determinant, we can perform the similarity transformation
\begin{equation}
\ket{\phi}\rightarrow e^{\frac{i}{2\hbar}\hat{p}^{2}}\ket{\phi},\quad\bra{\phi}\rightarrow\bra{\phi}e^{-\frac{i}{2\hbar}\hat{p}^{2}}.
\end{equation}
By using the formula
\begin{equation}
e^{-\frac{i}{2\hbar}\hat{p}^{2}}e^{-\frac{i}{2\hbar}\hat{x}^{2}}\kket p=\frac{1}{\sqrt{i}}e^{\frac{i}{2\hbar}p^{2}}\ket p,
\end{equation}
we obtain
\begin{align}
\mathcal{Z}_{k,N}^{\left[f\left(h_{{\rm D}}\right)\right]}= & i^{-\frac{N}{2}}e^{-\frac{i\pi\left(b^{2}+b^{-2}\right)}{12nk}\left(N^{3}-N\right)}\frac{1}{k^{N}}\hsum f\left(h_{{\rm D}}\right)\int\frac{d^{N}\phi}{N!}e^{-\sum_{s}^{N}\frac{i}{2\hbar}\left(n-1\right)h_{s}^{2}}\nonumber \\
 & \times\det\left(e^{\frac{2\pi i}{n}h_{t}\sigma_{s}}\bbraket{\frac{i}{bk}\sigma_{s}|\phi_{t}}\right)_{s,t}\det\left(e^{-\frac{2\pi i}{n}h_{s}\sigma_{t}}\braket{\phi_{s}|-\frac{ib}{k}\sigma_{t}}\right)_{s,t}.\label{eq:FGFcomp1}
\end{align}
We can diagonalize the second determinant by using
\begin{align}
 & \frac{1}{N!}\sum_{\left\{ h_{i}\right\} _{1\leq i\leq N}=0}^{n-1}f\left(h_{{\rm D}}\right)\int d^{N}\alpha\det\left(\left[a_{m}\left(\alpha_{n},h_{n}\right)\right]_{m,n}^{N\times N}\right)\det\left(\left[b_{n}\left(\alpha_{m},h_{m}\right)\right]_{m,n}^{N\times N}\right)\nonumber \\
 & =\sum_{\left\{ h_{i}\right\} _{1\leq i\leq N}=0}^{n-1}f\left(h_{{\rm D}}\right)\int d^{N}\alpha\det\left(\left[a_{m}\left(\alpha_{n},h_{n}\right)\right]_{m,n}^{N\times N}\right)\prod_{n}^{N}b_{n}\left(\alpha_{n},h_{n}\right),
\end{align}
and then perform the integration by using the delta functions\footnote{Strictly speaking, we should move the integration contour from $\mathbb{R}$ to $\mathbb{R}-\frac{ib}{k}\sigma_{t}$, so that the argument of the delta function becomes real.}
\begin{equation}
\braket{\phi_{s}|-\frac{ib}{k}\sigma_{t}}=\delta\left(\phi_{s}+\frac{ib}{k}\sigma_{t}\right),
\end{equation}
After decomposing the first determinant by using \eqref{eq:DetFormula} backwards, we obtain
\begin{align}
\mathcal{Z}_{k,N}^{\left[f\left(h_{{\rm D}}\right)\right]} & =e^{-i\theta\left(N,k\right)}\frac{1}{\left(nk\right)^{\frac{N}{2}}}\hsum f\left(h_{{\rm D}}\right)\nonumber \\
 & \quad\times e^{-\sum_{s}\frac{i\pi}{n}\left[k\left(n-1\right)h_{s}^{2}+2\sigma_{s}h_{s}\right]}\prod_{s<t}^{N}2\sin\frac{\pi}{n}\left(\frac{1}{k}\left(t-s\right)+\left(h_{t}-h_{s}\right)\right),
\end{align}
where
\begin{equation}
\theta\left(N,k\right)={\rm sgn}\left(k\right)\frac{\pi N^{2}}{4}+\frac{\pi\left(b^{2}+b^{-2}\right)}{12nk}\left(N^{3}-N\right).\label{eq:ThetaDef}
\end{equation}
The result including negative $k$ case is 
\begin{align}
\mathcal{Z}_{k,N}^{\left[f\left(h_{{\rm D}}\right)\right]}= & \frac{e^{-i\theta\left(N,k\right)}}{\left(n\left|k\right|\right)^{\frac{N}{2}}}\hsum f\left(h_{{\rm D}}\right)\nonumber \\
 & \times e^{-\sum_{s}^{N}\frac{i\pi}{n}\left[k\left(n-1\right)h_{s}^{2}+2{\rm sgn}\left(k\right)\sigma_{s}h_{s}\right]}\prod_{s<t}^{N}2\sin\frac{\pi}{n}\left(\frac{1}{\left|k\right|}\left(t-s\right)+\left(h_{t}-h_{s}\right)\right),\label{eq:FGFres}
\end{align}
This expression will be used for numerical study in section \ref{sec:Duality}.

In \eqref{eq:FGFres}, we have the explicit phase $e^{i\theta\left(N,k\right)}$, and the rest factor is real when $n=1$. However, when $n\geq2$, the rest part also has complex value. We study this point in section \ref{sec:Duality}.

Note that, when $f\left(h_{{\rm D}}\right)=1$, or equivalently, for $\mathcal{Z}_{k,N}^{\left[1\right]}=Z\left[{\rm U}\left(N\right)_{k,0}\right]$, we can obtain another simple expression. By using \eqref{eq:FGFres} and \eqref{eq:DetFormula}, we find
\begin{align}
\tilde{Z}\left[{\rm U}\left(N\right)_{k,0}\right]= & e^{-i\theta\left(N,k\right)}\frac{i^{-\frac{1}{2}N\left(N-1\right)}}{\left(n\left|k\right|\right)^{\frac{N}{2}}}\hsum\det\left(e^{-\frac{i\pi}{n}\left[k\left(n-1\right)h_{t}^{2}+2{\rm sgn}\left(k\right)\sigma_{t}h_{t}\right]}e^{\frac{2\pi i}{n}\sigma_{s}\left(\frac{t}{\left|k\right|}+h_{t}\right)}\right)_{s,t}.
\end{align}
The sum over $h_{i}$ can be put into the determinant
\begin{equation}
\tilde{Z}\left[{\rm U}\left(N\right)_{k,0}\right]=e^{-i\theta\left(k,N\right)}\frac{i^{-\frac{1}{2}N\left(N-1\right)}}{\left(n\left|k\right|\right)^{\frac{N}{2}}}\det\left(\sum_{h=0}^{n-1}e^{-\frac{i\pi}{n}\left[k\left(n-1\right)h^{2}+2{\rm sgn}\left(k\right)\sigma_{t}h\right]}e^{\frac{2\pi i}{n}\sigma_{s}\left(\frac{t}{\left|k\right|}+h\right)}\right)_{s,t}.
\end{equation}
In this expression, the multi-sum over $h_{i}$ becomes single sum.

\section{Level-rank dualities\label{sec:Duality}}

In this section, we study level-rank dualities. We will see that the matrix models match exactly across the dualities including non-trivial phases.

The strategy of identifying the phase is as follows. In section \ref{subsec:LRandPhase}, we identify the phase for the matrix model associated to ${\rm U}\left(N\right)_{k,0}$ gauge group using the well-known level-rank duality. We then identify the phase for the matrix model associated to ${\rm SU}\left(N\right)_{k}$ and ${\rm U}\left(N\right)_{k,k'}$ gauge group by using \eqref{eq:UtoSU} and \eqref{eq:w-woDiag}. In section \ref{subsec:LevelRankUU} and \ref{subsec:LevelRankUSU}, we show that this identification works for dualities studied in \cite{Hsin:2016blu}. This is non-trivial check of both the dualities and the phase.

Throughout this section, we use the exact values of the matrix models. The exact values are obtained by using \eqref{eq:FGFres} with \eqref{eq:UFGF} and \eqref{eq:SUFGF}.

\subsection{Phase factor from the well-known level-rank duality\label{subsec:LRandPhase}}

In this section, we study the phase using the well-known level-rank duality for pure Chern-Simons theory, which is the duality between ${\rm U}\left(N\right)_{k,0}$ and ${\rm U}\left(\left|k\right|-N\right)_{-k,0}$ gauge group \cite{Naculich:1990pa}. The values of the matrix models associated to ${\rm U}\left(N\right)_{k,0}$ and ${\rm U}\left(\left|k\right|-N\right)_{-k,0}$ gauge group are expected to be equal. However, we found that $\tilde{Z}\left[{\rm U}\left(N\right)_{k,0}\right]$ does not satisfy this relation.

To study in detail, we see the absolute value and the phase separately. 
\begin{table}
\begin{centering}
\begin{tabular}{|c||c|c||c|c||c|c|c|c|}
\hline 
$k$ & \multicolumn{2}{c||}{$3$} & \multicolumn{2}{c||}{$4$} & \multicolumn{4}{c|}{$5$}\tabularnewline
\hline 
$N$ & $1$ & $2$ & $1$ & $3$ & $1$ & $2$ & $3$ & $4$\tabularnewline
\hline 
\hline 
$n=1$ & $\frac{1}{\sqrt{3}}$ & $\frac{1}{\sqrt{3}}$ & $\frac{1}{2}$ & $\frac{1}{2}$ & $\frac{1}{\sqrt{5}}$ & $\frac{\sqrt{5-\sqrt{5}}}{5\sqrt{2}}$ & $\frac{\sqrt{5-\sqrt{5}}}{5\sqrt{2}}$ & $\frac{1}{\sqrt{5}}$\tabularnewline
\hline 
$n=2$ & $\frac{1}{\sqrt{3}}$ & $\frac{1}{\sqrt{3}}$ & $\frac{1}{\sqrt{2}}$ & $\frac{1}{\sqrt{2}}$ & $\frac{1}{\sqrt{5}}$ & $\frac{\sqrt{5+\sqrt{5}}}{5\sqrt{2}}$ & $\frac{\sqrt{5+\sqrt{5}}}{5\sqrt{2}}$ & $\frac{1}{\sqrt{5}}$\tabularnewline
\hline 
$n=3$ & $1$ & $1$ & $\frac{1}{2}$ & $\frac{1}{2}$ & $\frac{1}{\sqrt{5}}$ & $\frac{\sqrt{5+\sqrt{5}}}{5\sqrt{2}}$ & $\frac{\sqrt{5+\sqrt{5}}}{5\sqrt{2}}$ & $\frac{1}{\sqrt{5}}$\tabularnewline
\hline 
$n=4$ & $\frac{1}{\sqrt{3}}$ & $\frac{1}{\sqrt{3}}$ & $0$ & $0$ & $\frac{1}{\sqrt{5}}$ & $\frac{\sqrt{5-\sqrt{5}}}{5\sqrt{2}}$ & $\frac{\sqrt{5-\sqrt{5}}}{5\sqrt{2}}$ & $\frac{1}{\sqrt{5}}$\tabularnewline
\hline 
\end{tabular}
\par\end{centering}
\caption{\label{tab:CSabsNum}The exact values of $\left|\tilde{Z}\left[{\rm U}\left(N\right)_{k,0}\right]\right|$ for $3\protect\leq k\protect\leq5$, $1\protect\leq N\protect\leq k-1$ and $1\protect\leq n\protect\leq4$ with $b=1$. Only the values that are related to the duality are included here. The absolute values of the matrix models with Chern-Simons level $k$ and $-k$ are the same as in \eqref{eq:knMinus}. This table shows that the well-known level-rank duality holds for absolute value. We also numerically checked that $\left|\tilde{Z}\left[{\rm U}\left(k\right)_{k,0}\right]\right|=1$.}
\end{table}
The absolute values are in table \ref{tab:CSabsNum}. This table clearly shows that the duality holds for the absolute value. We expect that the duality holds including the phase.
\begin{table}
\begin{centering}
\begin{tabular}{|c|c||c|c||c|c|c||c|c|c|c|}
\hline 
$k$ & $2$ & \multicolumn{2}{c||}{$3$} & \multicolumn{3}{c||}{$4$} & \multicolumn{4}{c|}{$5$}\tabularnewline
\hline 
$N$ & $1$ & $1$ & $2$ & $1$ & $2$ & $3$ & $1$ & $2$ & $3$ & $4$\tabularnewline
\hline 
\hline 
$n=1$ & $-\frac{\pi}{4}$ & $-\frac{\pi}{4}$ & $\frac{2\pi}{3}$ & $-\frac{\pi}{4}$ & $\frac{3\pi}{4}$ & $\frac{3\pi}{4}$ & $-\frac{\pi}{4}$ & $\frac{4\pi}{5}$ & $\frac{19\pi}{20}$ & $0$\tabularnewline
\hline 
$n=2$ & $\setminus$ & $0$ & $-\frac{\pi}{6}$ & $-\frac{\pi}{4}$ & $-\frac{3\pi}{4}$ & $\frac{\pi}{4}$ & $-\frac{\pi}{2}$ & $\frac{9\pi}{10}$ & $-\frac{2\pi}{5}$ & $0$\tabularnewline
\hline 
$n=3$ & $\frac{\pi}{4}$ & $-\frac{\pi}{4}$ & $-\frac{13\pi}{18}$ & $-\frac{3\pi}{4}$ & $\frac{7\pi}{12}$ & $-\frac{5\pi}{12}$ & $\frac{\pi}{4}$ & $\frac{\pi}{15}$ & $\frac{19\pi}{60}$ & $-\frac{\pi}{3}$\tabularnewline
\hline 
$n=4$ & $0$ & $-\frac{\pi}{2}$ & $\frac{2\pi}{3}$ & $\setminus$ & $\setminus$ & $\setminus$ & $0$ & $\frac{\pi}{5}$ & $\frac{3\pi}{10}$ & $0$\tabularnewline
\hline 
\end{tabular}
\par\end{centering}
\caption{\label{tab:CSphaseNum}The exact values of the phases of $\tilde{Z}\left[{\rm U}\left(N\right)_{k,0}\right]$ with $b=1$. $\setminus$ means that the value of the matrix model is $0$. The value with negative $k$ can be obtained by taking the complex conjugate as in \eqref{eq:knMinus}. The duality between ${\rm U}\left(N\right)_{k,0}$ and ${\rm U}\left(\left|k\right|-N\right)_{-k,0}$ gauge group does not hold.}
\end{table}
However, the duality does not hold for the phase as in table \ref{tab:CSphaseNum}.

The phase already appeared in $n=1$ (squashed three-sphere) case, which is well studied. In this case, $\tilde{Z}\left[{\rm U}\left(N\right)_{k,0}\right]$ has the explicit phase $e^{-i\theta\left(N,k\right)}$ as in \eqref{eq:FGFres} with $f\left(h_{{\rm D}}\right)=1$ since the rest factors are explicitly real. In this case, this phase can be interpreted as the framing factor, and the framing factor should be removed when we consider the duality \cite{Marino:2004uf,Kapustin:2009kz}.
\begin{table}
\begin{centering}
\begin{tabular}{|c|c||c|c||c|c|c||c|c|c|c|}
\hline 
$k$ & $2$ & \multicolumn{2}{c||}{$3$} & \multicolumn{3}{c||}{$4$} & \multicolumn{4}{c|}{$5$}\tabularnewline
\hline 
$N$ & $1$ & $1$ & $2$ & $1$ & $2$ & $3$ & $1$ & $2$ & $3$ & $4$\tabularnewline
\hline 
\hline 
$n=1$ & $0$ & $0$ & $0$ & $0$ & $0$ & $0$ & $0$ & $0$ & $0$ & $0$\tabularnewline
\hline 
$n=2$ & $\setminus$ & $\frac{\pi}{4}$ & $\pi$ & $0$ & $\frac{3\pi}{8}$ & $\pi$ & $-\frac{\pi}{4}$ & $0$ & $\frac{\pi}{4}$ & $\pi$\tabularnewline
\hline 
$n=3$ & $\frac{\pi}{2}$ & $0$ & $\frac{7\pi}{18}$ & $-\frac{\pi}{2}$ & $-\frac{\pi}{3}$ & $\frac{\pi}{6}$ & $\frac{\pi}{2}$ & $-\frac{2\pi}{3}$ & $\frac{5\pi}{6}$ & $\frac{\pi}{3}$\tabularnewline
\hline 
$n=4$ & $\frac{\pi}{4}$ & $-\frac{\pi}{4}$ & $-\frac{\pi}{4}$ & $\setminus$ & $\setminus$ & $\setminus$ & $\frac{\pi}{4}$ & $-\frac{3\pi}{4}$ & $\frac{3\pi}{4}$ & $\frac{\pi}{2}$\tabularnewline
\hline 
\end{tabular}
\par\end{centering}
\caption{\label{tab:CSphaseNum2}The exact values of the phases of $e^{i\theta\left(N,k\right)}\tilde{Z}\left[{\rm U}\left(N\right)_{k,0}\right]$ with $b=1$. The duality between ${\rm U}\left(N\right)_{k,0}$ and ${\rm U}\left(\left|k\right|-N\right)_{-k,0}$ gauge group holds for $n=1$ case, but it does not hold for $n\protect\geq2$ case.}
\end{table}
After removing the framing factor, we find that the duality holds for $n=1$ case as in table \ref{tab:CSphaseNum2}.

On the other hand, as in table \ref{tab:CSphaseNum2}, the duality still does not hold for $n\geq2$ case. In this case, it is natural to expect that we need to generalize the phase \eqref{eq:ThetaDef}. After trials and errors, we found that \eqref{eq:AllPhaseDef} makes dualities complete. More explicitly, for ${\rm U}\left(N\right)_{k,0}$,
\begin{equation}
Z\left[{\rm U}\left(N\right)_{k,0}\right]=e^{i\Theta\left(N,k\right)}\tilde{Z}\left[{\rm U}\left(N\right)_{k,0}\right].\label{eq:PhasePFUdef}
\end{equation}
satisfies the expected relation of the duality
\begin{equation}
Z\left[{\rm U}\left(N\right)_{k,0}\right]=Z\left[{\rm U}\left(\left|k\right|-N\right)_{-k,0}\right].\label{eq:LRdualUU}
\end{equation}
Note that the $Z\left[{\rm U}\left(N\right)_{k,0}\right]$ is independent of the squashing parameter $b$. We checked this relation numerically by using \eqref{eq:UFGF} for various $N$, $k$ and $n$. 
\begin{table}
\begin{centering}
\begin{tabular}{|c|c||c|c||c|c|c||c|c|c|c|}
\hline 
$k$ & $2$ & \multicolumn{2}{c||}{$3$} & \multicolumn{3}{c||}{$4$} & \multicolumn{4}{c|}{$5$}\tabularnewline
\hline 
$N$ & $1$ & $1$ & $2$ & $1$ & $2$ & $3$ & $1$ & $2$ & $3$ & $4$\tabularnewline
\hline 
\hline 
$n=1$ & $0$ & $0$ & $0$ & $0$ & $0$ & $0$ & $0$ & $0$ & $0$ & $0$\tabularnewline
\hline 
$n=2$ & $0$ & $\frac{\pi}{2}$ & $-\frac{\pi}{2}$ & $\frac{\pi}{4}$ & $\pi$ & $-\frac{\pi}{4}$ & $0$ & $\frac{7\pi}{10}$ & $-\frac{7\pi}{10}$ & $0$\tabularnewline
\hline 
$n=3$ & $0$ & $\frac{\pi}{2}$ & $-\frac{\pi}{2}$ & $0$ & $\pi$ & $0$ & $\pi$ & $\frac{4\pi}{5}$ & $-\frac{4\pi}{5}$ & $\pi$\tabularnewline
\hline 
$n=4$ & $0$ & $\frac{\pi}{2}$ & $-\frac{\pi}{2}$ & $\setminus$ & $\setminus$ & $\setminus$ & $\pi$ & $-\frac{\pi}{2}$ & $\frac{\pi}{2}$ & $\pi$\tabularnewline
\hline 
\end{tabular}
\par\end{centering}
\caption{\label{tab:CSphaseDefNum}The exact values of the phases of $Z\left[{\rm U}\left(N\right)_{k,0}\right]$ with $b=1$. Only the values that are related to the duality are included here. The value with negative $k$ can be obtained by taking the complex conjugate as in \eqref{eq:knMinus}. This table shows that the duality \eqref{eq:LRdualUU} holds. We also numerically checked that $Z\left[{\rm U}\left(k\right)_{k,0}\right]$ is real.}
\end{table}
The part of the phase of $Z\left[{\rm U}\left(N\right)_{k,0}\right]$ is in table \ref{tab:CSphaseDefNum}.

For $n=1$ case, removing the framing factor is just equal to taking the absolute value. On the other hand, for $n\geq2$ case, the both side still have non-trivial phases as in table \ref{tab:CSphaseDefNum}.
In this sense, the check by using the lens space matrix models does something inherently more profound than simply checking that the absolute values are correct.

After identifying the correct phase for $Z\left[{\rm U}\left(N\right)_{k,0}\right]$, we can suppose the phase factor for $\tilde{Z}\left[{\rm U}\left(N\right)_{k,k'}\right]$ and $\tilde{Z}\left[{\rm SU}\left(N\right)_{k}\right]$ by using \eqref{eq:w-woDiag} and \eqref{eq:UtoSU}. We assume that these relations also hold for the matrix models with additional phase. Namely, we define the matrix models with

\begin{equation}
Z\left[{\rm U}\left(N\right)_{k,k'}\right]=\frac{Z\left[{\rm U}\left(1\right)_{kN+k'N^{2}}\right]}{Z\left[{\rm U}\left(1\right)_{kN}\right]}Z\left[{\rm U}\left(N\right)_{k,0}\right],\label{eq:w-woDiagP}
\end{equation}
and
\begin{equation}
Z\left[{\rm SU}\left(N\right)_{k}\right]=\frac{1}{N\times Z\left[{\rm U}\left(1\right)_{kN}\right]}Z\left[{\rm U}\left(N\right)_{k,0}\right].\label{eq:UtoSUP}
\end{equation}
The right hand side is already defined since all factors are written in terms of $Z\left[{\rm U}\left(N\right)_{k,0}\right]$. The explicit relations between before and after adding the phases, which are equivalent to \eqref{eq:w-woDiagP} and \eqref{eq:UtoSUP}, are in \eqref{eq:PhasePFdef}. Here we used the fact that the $k$ dependence of $\Theta\left(1,k\right)$ only comes from ${\rm sgn}\left(k\right)$ and $N\geq$1. Although \eqref{eq:w-woDiag} and \eqref{eq:UtoSU} holds only when $\gcd\left(N,n\right)=1$ and ${Z\left[{\rm U}\left(1\right)_{kN}\right]} \neq 0$, we found that \eqref{eq:PhasePFdef} works for any $N$ and $n$, which we will see in the following sections.

There are ambiguities in the phase \eqref{eq:AllPhaseDef} because some combinations of $N$, $k$ and $k'$ are invariant under the level-rank duality.
However, there is evidence to suggest that this choice is natural.
First, other choice of the phase generally breaks dualities including ${\rm U}\left(N\right)_{k,k'}$ or ${\rm SU}\left(N\right)_{k}$, which we study in the following sections, because the phase \eqref{eq:AllPhaseDef} appears in \eqref{eq:PhasePFdef} in a complex way.
Second, the phase \eqref{eq:AllPhaseDef} reduces to the known phase \eqref{eq:ThetaDef} when $n=1$.
Third, the phase \eqref{eq:AllPhaseDef} is very similar to the phase coming from the gravitational Chern-Simons term studied in \cite{Closset:2018ghr}.

\subsection{${\rm U}\left(N\right)_{k,-2{\rm sgn}\left(k\right)}$ and ${\rm U}\left(\left|k\right|-N\right)_{-k,2{\rm sgn}\left(k\right)}$\label{subsec:LevelRankUU}}

In this section, we study another level-rank duality proposed in \cite{Hsin:2016blu}, ${\rm U}\left(N\right)_{k,-2{\rm sgn}\left(k\right)}$ and ${\rm U}\left(\left|k\right|-N\right)_{-k,2{\rm sgn}\left(k\right)}$.\footnote{The explicit relation between our notation and the one of \cite{Hsin:2016blu} is
\begin{equation}
{\rm U}\left(N\right)_{k,k'}^{\text{[Our]}}={\rm U}\left(N\right)_{k-{\rm sgn}\left(k\right)N,k+k'N}^{\text{[HS]}},
\end{equation}
or, equivalently,
\begin{equation}
{\rm U}\left(N\right)_{K,K+NK'}^{\text{[HS]}}={\rm U}\left(N\right)_{K+{\rm sgn}\left(K\right)N,K'-{\rm sgn}\left(K\right)}^{\text{[Our]}}.
\end{equation}
} 

The expected relation is
\begin{equation}
Z\left[{\rm U}\left(N\right)_{k,-2{\rm sgn}\left(k\right)}\right]=Z\left[{\rm U}\left(\left|k\right|-N\right)_{-k,2{\rm sgn}\left(k\right)}\right].\label{eq:LRdualUU2}
\end{equation}
We checked this relation numerically by using \eqref{eq:UFGF} for various $N$, $k$ and $n$.
\begin{table}
\begin{centering}
\begin{tabular}{|c|c|c||c|c||c|c|c|c|}
\hline 
$k$ & \multicolumn{2}{c||}{$3$} & \multicolumn{2}{c||}{$4$} & \multicolumn{4}{c|}{$5$}\tabularnewline
\hline 
$N$ & $1$ & $2$ & $1$ & $3$ & $1$ & $2$ & $3$ & $4$\tabularnewline
\hline 
\hline 
$n=1$ & $1$ & $1$ & $\frac{1}{\sqrt{2}}$ & $\frac{1}{\sqrt{2}}$ & $\frac{1}{\sqrt{3}}$ & $\frac{\sqrt{5-\sqrt{5}}}{\sqrt{10}}$ & $\frac{\sqrt{5-\sqrt{5}}}{\sqrt{10}}$ & $\frac{1}{\sqrt{3}}$\tabularnewline
\hline 
$n=2$ & $1$ & $1$ & $0$ & $0$ & $\frac{i}{\sqrt{3}}$ & $\frac{\sqrt{5+\sqrt{5}}}{\sqrt{10}}e^{-\frac{3\pi i}{10}}$ & $\frac{\sqrt{5+\sqrt{5}}}{\sqrt{10}}e^{\frac{3\pi i}{10}}$ & $-\frac{i}{\sqrt{3}}$\tabularnewline
\hline 
$n=3$ & $1$ & $1$ & $-\frac{1}{\sqrt{2}}$ & $-\frac{1}{\sqrt{2}}$ & $i$ & $\frac{\sqrt{5+\sqrt{5}}}{\sqrt{10}}e^{-\frac{\pi i}{5}}$ & $\frac{\sqrt{5+\sqrt{5}}}{\sqrt{10}}e^{\frac{\pi i}{5}}$ & $-i$\tabularnewline
\hline 
$n=4$ & $1$ & $1$ & $-1$ & $-1$ & $\frac{i}{\sqrt{3}}$ & $-\frac{\sqrt{5-\sqrt{5}}}{\sqrt{10}}i$ & $\frac{\sqrt{5-\sqrt{5}}}{\sqrt{10}}i$ & $-\frac{i}{\sqrt{3}}$\tabularnewline
\hline 
\end{tabular}
\par\end{centering}
\caption{\label{tab:CSNumUU}The exact values of $Z\left[{\rm U}\left(N\right)_{k,-2}\right]$ for $3\protect\leq k\protect\leq5$, $1\protect\leq N\protect\leq k-1$ and $1\protect\leq n\protect\leq4$ with $b=1$. Only the values that are related to the duality are included here. The values of $Z\left[{\rm U}\left(N\right)_{-k,2}\right]$ can be obtained by taking the complex conjugate to $Z\left[{\rm U}\left(N\right)_{k,-2}\right]$ as in \eqref{eq:knMinus}. This table shows that the duality \eqref{eq:LRdualUU2} holds. We also numerically checked that $Z\left[{\rm U}\left(k\right)_{k,-2}\right]=1$. }
\end{table}
Table \ref{tab:CSNumUU} shows the exact values of $Z\left[{\rm U}\left(N\right)_{k,-2}\right]$ and shows that \eqref{eq:LRdualUU2} holds including the phase.

When $\gcd\left(N,n\right)=1$ and $\gcd\left(\left|k\right|-N,n\right)=1$, we can simplify \eqref{eq:LRdualUU2} by using \eqref{eq:w-woDiagP}. If we assume that the well-known level-rank duality \eqref{eq:LRdualUU} holds, we can further simplify it to
\begin{equation}
\frac{Z\left[{\rm U}\left(1\right)_{kN-2{\rm sgn}\left(k\right)N^{2}}\right]}{Z\left[{\rm U}\left(1\right)_{kN}\right]}=\frac{Z\left[{\rm U}\left(1\right)_{-k\left(\left|k\right|-N\right)+2{\rm sgn}\left(k\right)\left(\left|k\right|-N\right)^{2}}\right]}{Z\left[{\rm U}\left(1\right)_{-k\left(\left|k\right|-N\right)}\right]}.\label{eq:LRUU2-1}
\end{equation}
Because the matrix model associated to ${\rm U}\left(1\right)_{k}$ is factorized into the sum over $h_{i}$ and the integral over the Cartan, we can evaluate the integral independently. After performing the integration, we find
\begin{align}
Z\left[{\rm U}\left(1\right)_{k}\right] & =\frac{1}{\sqrt{\left|k\right|}}H\left(k\right),\label{eq:U1Decom}
\end{align}
where
\begin{equation}
H\left(k\right)=\frac{i^{{\rm sgn}\left(k\right)\frac{n-1}{2}}}{\sqrt{n}}\sum_{h=0}^{n-1}e^{-\frac{i\pi k}{n}\left(n-1\right)h^{2}}.
\end{equation}
Using this result, \eqref{eq:LRUU2-1} further reduces to
\begin{equation}
H\left(-k\left(\left|k\right|-N\right)\right)H \left(kN-2{\rm sgn}\left(k\right)N^{2}\right)=H\left(kN\right) H\left(-k\left(\left|k\right|-N\right)+2{\rm sgn}\left(k\right)\left(\left|k\right|-N\right)^{2}\right).\label{eq:eqG1}
\end{equation}
Since this expression is simpler than \eqref{eq:LRdualUU2}, we checked this relation up to higher $k$, $N$ and $n$.

\subsection{${\rm U}\left(N\right)_{k,-{\rm sgn}\left(k\right)}$ and ${\rm SU}\left(\left|k\right|-N\right)_{-k}$\label{subsec:LevelRankUSU}}

In this section, we study the level-rank duality between ${\rm U}\left(N\right)_{k,-{\rm sgn}\left(k\right)}$ and ${\rm SU}\left(\left|k\right|-N\right)_{-k}$ proposed in \cite{Hsin:2016blu}. 

The expected relation is
\begin{equation}
Z\left[{\rm U}\left(N\right)_{k,-{\rm sgn}\left(k\right)}\right]=Z\left[{\rm SU}\left(\left|k\right|-N\right)_{-k}\right].\label{eq:LRdualUSU}
\end{equation}
We checked this relation numerically by using \eqref{eq:SUFGF} for various $k$, $N$ and $n$.
\begin{table}
\begin{centering}
\begin{tabular}{|c|c|c|c||c|c|c||c|c|c|c|}
\hline 
$k$ & $2$ & \multicolumn{2}{c||}{$3$} & \multicolumn{3}{c|}{$4$} & \multicolumn{4}{c|}{$5$}\tabularnewline
\hline 
$N$ & $1$ & $1$ & $2$ & $1$ & $2$ & $3$ & $1$ & $2$ & $3$ & $4$\tabularnewline
\hline 
\hline 
$n=1$ & $1$ & $\frac{1}{\sqrt{2}}$ & $1$ & $\frac{1}{\sqrt{3}}$ & $\frac{1}{2}$ & $1$ & $\frac{1}{2}$ & $\frac{\sqrt{5-\sqrt{5}}}{\sqrt{30}}$ & $\frac{\sqrt{5-\sqrt{5}}}{2\sqrt{5}}$ & $1$\tabularnewline
\hline 
$n=2$ & $1$ & $0$ & $1$ & $\frac{i}{\sqrt{3}}$ & $\frac{\sqrt{2-\sqrt{2}}}{2}e^{-\frac{3\pi i}{8}}$ & $1$ & $\frac{1}{\sqrt{2}}e^{\frac{\pi i}{4}}$ & $\frac{\sqrt{5+\sqrt{5}}}{\sqrt{30}}e^{-\frac{4\pi i}{5}}$ & $0$ & $1$\tabularnewline
\hline 
$n=3$ & $1$ & $-\frac{1}{\sqrt{2}}$ & $1$ & $i$ & $\frac{1}{2}$ & $1$ & $\frac{1}{2}$ & $\frac{\sqrt{5+\sqrt{5}}}{\sqrt{10}}e^{-\frac{7\pi i}{10}}$ & $\frac{\sqrt{5+\sqrt{5}}}{2\sqrt{5}}e^{-\frac{4\pi i}{5}}$ & $1$\tabularnewline
\hline 
$n=4$ & $1$ & $-1$ & $1$ & $\frac{i}{\sqrt{3}}$ & $\frac{1}{\sqrt{2}}e^{\frac{\pi i}{4}}$ & $1$ & $0$ & $-\frac{\sqrt{5-\sqrt{5}}}{\sqrt{30}}$ & $-\frac{\sqrt{5-\sqrt{5}}}{\sqrt{10}}i$ & $1$\tabularnewline
\hline 
\end{tabular}
\par\end{centering}
\caption{\label{tab:CSabsNumUSU}The exact values of the both side of \eqref{eq:LRdualUSU} for $2\protect\leq k\protect\leq5$, $1\protect\leq N\protect\leq k-1$ and $1\protect\leq n\protect\leq4$ with $b=1$.}
\end{table}
Table \ref{tab:CSabsNumUSU} shows the exact values of the both side of \eqref{eq:LRdualUSU}.

When $\gcd\left(N,n\right)=1$ and $\gcd\left(\left|k\right|-N,n\right)=1$, we can simplify \eqref{eq:LRdualUSU} by using \eqref{eq:w-woDiagP} and \eqref{eq:UtoSUP}. If we assume that the well-known level-rank duality \eqref{eq:LRdualUU} holds, it further reduces to
\begin{equation}
\frac{Z\left[{\rm U}\left(1\right)_{kN-{\rm sgn}\left(k\right)N^{2}}\right]}{Z\left[{\rm U}\left(1\right)_{kN}\right]}=\frac{1}{\left(\left|k\right|-N\right)Z\left[{\rm U}\left(1\right)_{-k\left(\left|k\right|-N\right)}\right]}.
\end{equation}
After performing the integration of the Cartan, we find
\begin{equation}
H\left(-k\left(\left|k\right|-N\right)\right) H\left(kN-{\rm sgn}\left(k\right)N^{2}\right)=H\left(kN\right).
\end{equation}
Since this expression is simpler than \eqref{eq:LRdualUSU}, we checked this relation up to higher $k$, $N$ and $n$.


\section{Discussion}
\label{Discussion}

We have investigated a way to compute topological phase in three dimensional supersymmetric field theory including topological field theory. 
To this end we have developed a Pauli-Villars regularization preserving supersymmetry in the supersymmetric localization by applying to the computation of partition function for pure Chern-Simons theory on lens space. 
We have computed the background dependent phase factor coming from the Chern-Simons term and reproduced the result in \cite{Imamura:2012rq,Imamura:2013qxa} rather than in \cite{2002math......9403H,Brini:2008ik,Gang:2009qdj,Alday:2013lba}. 
We have tested the resulting partition function whether to exhibit level rank dualities including the one recently proposed in \cite{Hsin:2016blu} and confirmed the precise agreement of the absolute value for all ranks and levels within our numerical calculation. 
We have argued that this partition function depends on a correctly chosen spin structure for the duality to hold.  
We have also shown that there exists a phase factor with which the lens space partition function enjoys the perfect match between any level rank dual pair including the overall phase. 
This provides strong evidence for the level rank dualities to hold including spin Chern-Simons theory.  

In the procedure given in \cite{Witten:1988hf} the topological phase was determined by adding a gravitational Chern-Simons term as a finite counter term.
In the supersymmetric localization and regularization it will be necessary to add a supersymmetrized gravitational Chern-Simons term to obtain a correct topological phase. We expect that a phase factor for the level rank dualities to hold including the total phase can be reproduced from a correct counter term. 
(See \cite{Closset:2012ru,Imbimbo:2014pla,Closset:2018ghr} for relevant references.)

It would be interesting to compute the presented phase factor for the duality from the corresponding two dimensional WZNW theory. 
For this goal one needs to recompute modular matrices for the WZNW theory to be spin  by generalizing the result in the abelian case \cite{Okuda:2020fyl}. 
The correct generalization will include the effect for the addition of suitable background fields to Chern-Simons matter theory argued in \cite{Hsin:2016blu}.  

An important observation achieved in this paper is that the $SU(N)_k$ partition function on lens space, which enjoys the level/rank duality, has the similar deviation in terms of the relative phase as in the case of $U(N)$. It is known that the $SU(N)_k$ WZNW model does not depend on the spin structure, and suitable background fields need to be turned on in order to achieve the level/rank duality \cite{Hsin:2016blu}. Therefore the deviation of the relative phase does not originate in spin structure but in a certain background  field. We presume that this is a background gauge field turned on in order to preserve supersymmetry. 

It is also interesting whether the phase factor presented in this paper can be obtained in other computation techniques such that the perturbation or the surgery. In that case, we have to be care of the framing because the framing used in the supersymmetric localization technique is different from the canonical one. (Notice that the phase factor we find reduces to the framing factor when the manifold is $S_{b}^{3}$, or equivalently, $n=0$.) It would be nice to study the phase factor associated to manifolds where the supersymmetric technique cannot be used by using these computation techniques.

It is known that pure Chern-Simons theory can be embedded into string theory by using D-branes and the level rank duality becomes manifest in the context of AdS/CFT duality from the brane configuration 
\cite{Fujita:2009kw,Armoni:2014cia}. 
It would be interesting to study whether the topological phase studied in this paper plays any role in this context.  

We hope to come back to these issues in near future.

\section*{Acknowledgments}
We would like to thank Takuya Okuda for participation up to a middle stage of the project and for valuable discussions.
We are also grateful to Keita Nii for valuable discussions and comments.
NK is supported by Grant-in-Aid for JSPS Fellows No.20J12263.
SY is supported in part by that of the Japanese Ministry of Education, Sciences
and Technology, Sports and Culture (MEXT) for Scientific Research (No. JP19K03847).

\appendix 
\section{Computation of gravitational Chern-Simons term} 
\label{gcs} 

In this appendix we compute the gravitational Chern-Simons term for the squashed lens spaces. 

First we consider the squashed lens space $L^{(1)}_b(n,1)$ obtained from the squashed three sphere whose metric is $ds^2 =\ell_1^2 + \ell_2^2 + \frac1{v^2}\ell_3^2$, 
where $\ell_i$ is the left-invariant 1-form defined by 
\beal{
\ell_1 =& \frac{1}{2} (d\theta \sin \psi-d\phi \sin \theta \cos \psi) , ~~ 
\ell_2 =\frac{1}{2} (d\theta \cos \psi+d\phi \sin \theta \sin \psi), ~~ 
\ell_3 =\frac{1}{2} (d\psi+d\phi \cos \theta),
}
where $0 \leq \theta \leq \pi, 0 \leq \phi < 2\pi, 0 \leq \psi < \frac{4\pi}{|n|}$.  
An orthonormal frame is $e^i= v^{-\delta_{i,3}}\ell_i$. 
The connection 1-form is computed as 
\beal{
\omega^{12}
=\frac{\left(1-2 v^2\right) }{v^2}\ell_3, ~~
\omega^{23}
=-\frac1v \ell_1, ~~ 
\omega^{31}
=-\frac1v \ell_2, ~~
}
from which one can compute the gravitational Chern-Simons form as 
\be 
\tr [\omega \mr d\omega +\frac23 \omega\wedge\omega\wedge \omega]  
=\frac{1-2v^2+2v^4} {v^4} \sin\theta \mathrm{d}\theta \wedge \mathrm{d}\phi \wedge \mathrm{d}\psi . 
\ee
As a result the gravitational Chern-Simons term evaluates to 
\be 
I[L^{(1)}_b(n,1)] = \frac1{4\pi} \int_{L_b(n,1)} \tr [\omega \mr d\omega +\frac23 \omega^{3}] 
=\frac{b^8-4 b^6+22 b^4-4 b^2+1}{16 b^4} \frac{4\pi}n, 
\ee
where we used $v=\frac{2b}{1+b^2}$.

Next we consider the squashed lens space $L^{(2)}_b(n,1)$ obtained from the squashed three sphere given as $b^2(x_1^2+x_2^2)+ b^{-2}(x_3^2+x_4^2)=1.$ by taking the $\mbb Z_{n}$ quotient. 
We use parametrization such that 
\beal{
\pmat{x_1 \\ x_2 \\ x_3 \\ x_4 }
= \pmat{b^{-1} \cos\frac\theta2 \cos\frac{\phi+\psi}2 \\ - b^{-1} \cos\frac\theta2 \sin\frac{\phi+\psi}2 \\ b \sin\frac\theta2 \cos\frac{\phi-\psi}2 \\ b \sin\frac\theta2 \sin\frac{\phi-\psi}2   }.
}
Then the metric is given by 
\beal{
ds^2 
=&\frac14\{  f(\theta)^2 d\theta^2 + \frac{s(\theta)^2 }{f(\theta)^2 } (d\phi)^2 + f(\theta)^2 (d\psi + \frac{c(\theta)^2}{f(\theta)^2}d\phi)^2 \} \notag
}
where 
\beal{ 
c(\theta)=&\sqrt{b^{-2}\cos^2(\frac\theta2) -b^2 \sin^2(\frac\theta2)}, \nn
s(\theta)=&\sqrt{(b^{-2} +b^2\cos\theta)(b^2 - b^{-2}\cos\theta)}, \nn
f(\theta) =& \sqrt{b^{-2} \sin^2(\frac\theta2) + b^2 \cos^2(\frac\theta2)}.
}

A natural choice of orthnormal frame may be 
\beal{
e^{\hat\theta}=& \frac{f(\theta)}2   d\theta, ~~
e^{\hat\phi}= \frac{s(\theta) }{2f(\theta) }d\phi, ~~
e^{\hat\psi}=\frac{f(\theta)}2 (d\psi + \frac{c(\theta)^2}{f(\theta)^2}d\phi), ~~\notag
}
The connection 1-form is computed as 
\beal{
\omega^{\hat\psi\hat\phi}
=&\frac{\left(b^{-4}+b^4\right) \sin \theta }{4 f(\theta)^2 s(\theta) }\mathrm{d}\theta , \nn
\omega^{\hat\theta\hat\phi}
=&-\frac{\sin \theta  \left(b^{-2} (\cos \theta -1)-b^2 (\cos \theta +1)\right){}^2 \left(2  \mathrm{d}\phi  \cos \theta +b^{-4} (\mathrm{d}\phi -\mathrm{d}\psi )-b^4 (\mathrm{d}\psi +\mathrm{d}\phi )\right)}{16 f(\theta)^6 s(\theta)}, \nn 
\omega^{\hat\theta\hat\psi}
=& \frac{\sin \theta  \left(b^{-2} (\mathrm{d}\phi -\mathrm{d}\psi )+b^2 (\mathrm{d}\psi +\mathrm{d}\phi )\right)}{4f(\theta)^2}. 
}
From this one can compute the gravitational Chern-Simons form, which vanishes before performing the volume integration: 
\be 
\tr [\omega \mr d\omega +\frac23 \omega\wedge\omega\wedge \omega]  = 0. 
\ee

On the other hand, we choose another frame which reduces to the set of the left-invariant 1-form in the round sphere limit $b\to1$: 
\beal{
e^1= \sin\psi e^{\hat\theta} -\cos\psi e^{\hat\phi}, ~~
e^2= \cos\psi e^{\hat\theta} +\sin\psi e^{\hat\phi}, ~~
e^3 = e^{\hat\psi}.
}
The connection 1-form is computed as 
\beal{
\omega^{21}
=&-A\left(b^{-4} \sin \theta  (\mathrm{d}\phi -\mathrm{d}\psi )-b^4 \sin \theta  (\mathrm{d}\psi +\mathrm{d}\phi )+4 ( b^2 \cos ^2(\frac{\theta }{2})+ b^{-2} \sin^2(\frac{\theta }{2})) s(\theta) \mathrm{d}\psi + \sin (2 \theta ) \mathrm{d}\phi  \right) , \nn
\omega^{31}
=&A\sin \theta \left((b^4+b^{-4}) \cos \psi \mathrm{d}\theta  +s(\theta) \sin \psi \{ b^{-2}(\mathrm{d}\phi -\mathrm{d}\psi ) +b^2  (\mathrm{d}\psi +\mathrm{d}\phi ) \} \right), \nn 
\omega^{32}
=&-A\sin \theta \left( (b^4+b^{-4})  \sin \psi \mathrm{d}\theta -s(\theta)\cos \psi  \{ b^{-2} (\mathrm{d}\phi -\mathrm{d}\psi )  -b^2  (\mathrm{d}\psi +\mathrm{d}\phi ) \} \right),
}
where 
\be 
A= \frac{ \left(b^{-2} (\cos \theta -1)-b^2 (\cos \theta +1)\right) }{8 s(\theta) f(\theta)^4}. 
\ee
The gravitational Chern-Simons form is computed as 
\beal{
\tr [\omega \mr d\omega +\frac23 \omega^{3}] 
=& \frac{\left(b^{-2}+b^2\right) \left(b^{-2} (\cos \theta-1)-b^2 (\cos \theta+1)\right){}^4 s(\theta) }{512 \left(b^2-b^{-2} \cos \theta\right){}^2 \left(b^2 \cos \theta+b^{-2}\right){}^2 \left(b^{-2} \sin ^2\left(\frac{\theta }{2}\right)+b^2 \cos ^2\left(\frac{\theta }{2}\right)\right){}^6} \nn
&\times \bigg(b^{-8} (-9 \cos \theta+2 \cos (2 \theta )+\cos (3 \theta )+6)+4 b^{-4} (7 \cos \theta-3 \cos (2 \theta )+\cos (3 \theta )-1) \nn
&+2  (6 \cos (2 \theta )+\cos (4 \theta )+1)-4 b^4(7 \cos \theta+3 \cos (2 \theta )+\cos (3 \theta )+1) \nn
&+b^8 (9 \cos \theta+2 \cos (2 \theta )-\cos (3 \theta )+6)\bigg) \mathrm{d}\theta \wedge \mathrm{d}\phi \wedge \mathrm{d}\psi . 
}
From this the gravitational Chern-Simons term evaluates to
\be 
I[L^{(2)}_b(n,1)] = \frac1{4\pi} \int_{L_b(n,1)} \tr [\omega \mr d\omega +\frac23 \omega^{3}] =0 ,
\ee
because the indefinite integral of the Chern-Simons form is computed as 
\beal{
\int \tr [\omega \mr d\omega +\frac23 \omega^{3}] 
=\frac{\sin \theta \left(2  \cos \theta+b^{-4}-b^4\right) }{s(\theta) \left(b^{-2} (\cos \theta-1)-b^2 (\cos \theta+1)\right)}. 
}
This result is not inconsistent with the round sphere limit, since in the limit $b\to1$ the gravitational Chern-Simons form reduces to 
\beal{
\tr [\omega \mr d\omega +\frac23 \omega^{3}] 
\to  \sin\theta \mathrm{d}\theta \wedge \mathrm{d}\phi \wedge \mathrm{d}\psi . 
}
As a result the value of the integration changes as  $I[L_1(n,1)] =  \frac{4\pi}n.$
In particular, $I[L_1(n,1)]=\lim_{b\to1} I[L^{(1)}_b(n,1)] \not= \lim_{b\to1} I[L^{(2)}_b(n,1)]$. 

\section{$\eta$ invariant on lens space with holonomy}
\label{eta-invariant} 
In this appendix we compute the eta invariant on lens space under non-trivial holonomy background. For computational simplicity we leave a lens space unsquashed.

The eta invariant is computed by the zero limit of the eta series defined by
\be
\eta(z) := \Tr [ \sgn(\sla D) |\sla D|^{-z} ], \quad 
\ee
where $\sla D$ is the Dirac operator on lens space with holonomy background, which may be referred to as the twisted Dirac operator. 

The eta series can be computed once we know the eigenvalues and its degeneracy of the twisted Dirac operator. The eigenfunctions of the Dirac operator are given by the spinor spherical harmonics, which are in the $(\frac k2,\frac{k+1}2)$ or $(\frac{k+1}2,\frac k2)$ representations for the global symmetry of the three sphere, $SU(2)_L\times SU(2)_R$. The former has the eigenvalue $-(k+\frac32)$. while the latter $(k+\frac32)$. $j_3$ and $\wt j_3$ denote the weight for the representation or the angular momentum for $SU(2)_L$ and $SU(2)_R$, respectively. 
Then the effect of the holonomy $h$ is implemented as the truncation of the modes such that there exists an integer $l$ to restrict $\wt j_3$ as 
\be 
\wt j_3 = {h + l n \over 2} .
\label{twistSelection}
\ee
Notice that this shows that the result is invariant under the replacement $h \to n-h$ so that we can assume $h\leq \frac n2$ without losing generality. 
These data enable us to compute the eta series for the twisted Dirac operator as 
\beal{
\eta(z)
=&\sum_{k\in\mbb N} \{ \sum_{\wt j_3 \in \{-\frac{k+1}2, \cdots, \frac{k+1}2\} \atop 2\wt j_3 -h \in n\mbb Z} \sum_{ j_3 =-\frac k2 }^{\frac k2}- (k + {3\over 2})^{-z}+ \sum_{\wt j_3 \in \{- \frac k2, \cdots, \frac k2\} \atop 2\wt j_3 -h \in n\mbb Z} \sum_{ j_3 =-(\frac{k+1}2) }^{\frac{k+1}2}+(k + {3\over 2})^{-z} \} 
\nn
=&\sum_{k=0}^\infty \{-(k+1) N^{(n,h)}_{k+1} (k + {3\over 2})^{-z} + (k+2) N^{(n,h)}_{k} (k + {3\over 2})^{-z} \},
}
where $N^{(n,h)}_{k}$ is defined by the number of the elements of the set $\{ l \in \mbb Z |  {h + l n \over 2} = -{k\over 2} + m ,~ \exists m=0,1, \cdots, k \} $.
The eta series can be written as 
\beal{
\eta(z)
=&- f(z-1;\half) +\half  f(z;\half) + f(z-1;{3\over 2}) + \half f(z;{3\over 2}) ,
}
where 
\beal{
f(s;a)=& \sum_{k=0}^\infty N^{(n,h)}_{k} (k + a)^{-s} .
}

In what follows, we compute this separating the cases whether $n$ is even or odd. To this end we use notation such that the quotient and reminder of $x$ divided by $m$ by $(x)_m$ and $[x]_m$ respectively.
We first consider that $n$ is even: $n=2p$. 
In this case $N^{(2p,h)}_{k}$ does not vanish if and only if $k+h$ is even. Indeed we can compute it as  
\beal{ 
N^{(2p,h)}_{k} 
=&\left\{ 
\begin{array}{lllll} 
2(r)_p  & ([r]_p <{h-[h]_2\over2} ) \\
2(r)_p +1  & ({h-[h]_2\over2} \leq [r]_p <p - {h+[h]_2\over2}  ) \\
2(r)_p +2  & (p - {h+[h]_2\over2}\leq [r]_p     )   \\
\end{array} \right. ,
}
where $r$ is defined by $k= 2r+[h]_2$. 
Employing this we can compute $f(s;a)$ as 
\beal{
f(s;a) 
=& \sum_{r=0}^\infty N^{(2p,h)}_{2r+[h]_2} (2r+[h]_2 + a)^{-s} \nn
=&\sum_{(r)_p=0}^\infty \bigg(  \sum_{ [r]_p = 0}^{{h-[h]_2\over2}-1} { 2(r)_p  \over  (2(r)_p p +2 [r]_p+[h]_2 + a)^{s} } + \sum_{ [r]_p = {h-[h]_2\over2} }^{p-{h+[h]_2\over 2}-1} { 2(r)_p +1 \over  (2(r)_p p + 2[r]_p+[h]_2 + a)^{s} } \nn
&+ \sum_{ [r]_p = p-{h+[h]_2\over 2}}^{p-1} { 2(r)_p + 2 \over  (2(r)_p p + 2[r]_p+[h]_2 + a)^{s} } \bigg) \nn
=&{1\over (2p)^s} \bigg( 2 \sum_{ [r]_p = 0}^{p-1} \{ \zeta(s-1; {2[r]_p +[h]_2+ a \over 2p}) - {2[r]_p +[h]_2+ a \over 2p} \zeta(s; {2[r]_p +[h]_2+ a \over 2p}) \} \nn
&+\sum_{ [r]_p = {h-[h]_2\over2} }^{p-{h+[h]_2\over 2}-1}  \zeta(s; {2[r]_p +[h]_2+ a \over 2p}) +2 \sum_{ [r]_p = p-{h+[h]_2\over 2}}^{p-1} \zeta(s; {2[r]_p +[h]_2+ a \over 2p}) \bigg) ,
}
where we used the Hurwitz zeta function $
\zeta(s;a) = \sum_{n=0}^\infty {1\over (n+a)^s}.$
Therefore the $\eta$ invariant for the twisted Dirac operator is computed as follows. 
\beal{
\eta(L_1(n,1);h)
=&- f(-1;\half) +\half  f(0;\half) + f(-1;{3\over 2}) +\half f(0;{3\over2}) \nn
=&\frac{6 h^2-6 h n+n^2-1}{6 n},
}
where we used the zeta function regularization $
\zeta(0;a) = \half -a , ~ \zeta(-1;a) = - { a^2 - a + 1/6 \over 2}$.

Next we consider the case with $n$ odd.  
In this case $N^{(n,h)}_{k} $ does not vanish for any value of $k+h$. 
In fact, when $k+h$ is odd, it is computed as 
\beal{ 
N^{(n,h)}_{k} 
=&\left\{ 
\begin{array}{lllll} 
2(r)_n  & ([r]_n <{n- h+[h]_2-1\over2} ) \\
2(r)_n +1 & ({n- h+[h]_2-1\over2}\leq [r]_n < {n+h+[h]_2-1\over2} ) \\
2(r)_n +2  & ( {n+h+[h]_2-1\over2}\leq [r]_n    )   \\
\end{array} \right. ,
}
where $r$ is defined by $k= 2r+{1-[h]_2}$.
When $k+h$ is even, it is computed as 
\beal{ 
N^{(n,h)}_{k} 
=&\left\{ 
\begin{array}{lllll} 
2(r)_n  & ([r]_n <{h-[h]_2\over2} ) \\
2(r)_n +1 & ({h-[h]_2\over2} \leq [r]_n < n - {h+ [h]_2\over2} ) \\
2(r)_n +2  & ( n - {h+ [h]_2\over2}\leq [r]_n    )   \\
\end{array} \right. ,
}
where we set $k= 2r+[h]_2$. 
Using this we can compute $f(s;a)$ as 
\beal{
f(s;a) =& \sum_{r=0}^\infty ( N^{(n,h)}_{2r+1-[h]_2} (2r+1-[h]_2 + a)^{-s} + N^{(n,h)}_{2r+[h]_2} (2r+[h]_2 + a)^{-s} ).
}
where the first summation is further computed as 
\beal{
&\sum_{r=0}^\infty  N^{(n,h)}_{2r+1-[h]_2} (2r+1-[h]_2 + a)^{-s} \nn
=&{1\over (2n)^s} \bigg( 2 \sum_{ [r]_n = 0}^{n-1} \{ \zeta(s-1; {2[r]_n +1-[h]_2+ a \over 2n}) - {2[r]_n +1-[h]_2+ a \over 2n} \zeta(s; {2[r]_n +1-[h]_2+ a \over 2n}) \} \nn
&+ \sum_{ [r]_n = {n-h+[h]_2-1\over2} }^{n-1}  \zeta(s; {2[r]_n +1-[h]_2+ a \over 2n}) + \sum_{ [r]_n ={n+h+[h]_2-1\over2}}^{n-1}  \zeta(s; {2[r]_n +1-[h]_2+ a \over 2n}) \bigg),  \notag 
}
while the second one is 
\beal{
&\sum_{r=0}^\infty ( N^{(n,h)}_{2r+[h]_2} (2r+[h]_2 + a)^{-s} ) \nn
=&{1\over (2n)^s} \bigg( 2 \sum_{ [r]_n = 0}^{n-1} \{ \zeta(s-1; {2[r]_n +[h]_2+ a \over 2n}) - {2[r]_n +[h]_2+ a \over 2n} \zeta(s; {2[r]_n +[h]_2+ a \over 2n}) \} \nn
&+\sum_{ [r]_n = {h-[h]_2\over2} }^{n -1}  \zeta(s; {2[r]_n +[h]_2+ a \over 2n}) +\sum_{ [r]_n = n-{h+[h]_2\over 2}}^{n-1} \zeta(s; {2[r]_n +[h]_2+ a \over 2n}) \bigg) . 
}
From this expression the $\eta$ invariant is computed as 
\beal{
\eta(L_1(n,1);h)
=&\frac{6 h^2-6 h n+n^2-1}{6 n}. 
}
In both cases, we obtain the same expression and reproduce the result in \cite{CisnerosMolina2001TheO}.

\section{Decomposition of the sum over $h_{i}$\label{sec:DecomLemma}}

In this appendix we show a formula, which is useful for decomposing the holonomy sum. 

We assume that positive integers $N$ (which denotes the rank in the main text) and $n$ (which is the parameter of the geometry $L_{b}\left(n,1\right)$ in the main text) satisfy $\gcd\left(N,n\right)=1$. We also assume that a function $f\left(\vec{h}\right)$ satisfies $f\left(\vec{h}\right)=f\left(\vec{h}+n\vec{1}\right)$, where $\vec{1}=\left(1,1,\ldots,1\right)$. The formula
\begin{equation}
\hsum f\left(\vec{h}\right)=\sum_{\tilde{h}=0}^{n-1}\hsum\delta_{h_{{\rm D}},0}^{{\rm mod}\thinspace n}f\left(\vec{h}+\tilde{h}\vec{1}\right),\label{eq:DecomLemma}
\end{equation}
holds, where $h_{{\rm D}}=\sum_{s}^{N}h_{s}$, and the definition of $\delta_{h_{{\rm D}},0}^{{\rm mod}\thinspace n}$ is in \eqref{eq:ModDeltaDef}.

In the following, we show this formula.
First, the left hand side of \eqref{eq:DecomLemma} becomes
\begin{equation}
\hsum f\left(\vec{h}\right)=\sum_{\tilde{h}=0}^{n-1}\hsum\delta_{h_{{\rm D}},\tilde{h}}^{{\rm mod}\thinspace n}f\left(\vec{h}\right).
\end{equation}
Second, we use the following fact. Since $N,n\in\mathbb{N}$ and $\gcd\left(N,n\right)=1$, for any $g\in\mathbb{Z}$, there exists $c_{g}\in\mathbb{Z}$ such that 
\begin{equation}
g-c_{g}N\in n\mathbb{Z}.\label{eq:cProp1}
\end{equation}
Furthermore, we can chose $c_{g}$ as 
\begin{equation}
\left\{ c_{g}|g=0,1,\ldots,n-1\right\} =\left\{ 0,1,\ldots,n-1\right\} .\label{eq:cProp2}
\end{equation}
Since $f\left(\vec{h}\right)=f\left(\vec{h}+n\vec{1}\right)$ and $\delta_{h_{{\rm D}},\tilde{h}}^{{\rm mod}\thinspace n}=\delta_{\sum_{s}^{N}\left(h_{s}+n\right),\tilde{h}}^{{\rm mod}\thinspace n}$, we can shift the holonomy sum $\hsum$ to $\sum_{h_{1},\ldots,h_{N}=c_{\tilde{h}}}^{n-1+c_{\tilde{h}}}$. Therefore,
\begin{equation}
\hsum f\left(\vec{h}\right)=\sum_{\tilde{h}=0}^{n-1}\hsum\delta_{\sum_{s}^{N}\left(h_{s}+c_{\tilde{h}}\right),\tilde{h}}^{{\rm mod}\thinspace n}f\left(\vec{h}+c_{\tilde{h}}\vec{1}\right).
\end{equation}
Since we chose $c_{\tilde{h}}$ so that \eqref{eq:cProp1} and \eqref{eq:cProp2} hold, we finally obtain \eqref{eq:DecomLemma}.

\bibliographystyle{utphys}
\bibliography{levelrank}

\end{document}